\documentclass[prl,reprint,nofootinbib, preprintnumbers, floatfix, unsortedaddress, superscriptaddress, longbibliography,times]{revtex4-2}

\usepackage{orcidlink}
\usepackage{multirow}
\usepackage{siunitx}
\usepackage{colortbl}
\usepackage{dcolumn}

\usepackage{tikz}
\usetikzlibrary{decorations.markings, arrows, decorations.pathmorphing}

\usepackage{graphicx}
\usepackage{latexsym}
\usepackage{amsmath,amssymb}
\usepackage{amsfonts}
\usepackage{bm}
\usepackage{bbm}
\usepackage{microtype}
\usepackage{xspace}
\usepackage{placeins}

\usepackage{hyperref}
\usepackage[capitalise]{cleveref}
\usepackage{braket}
\usepackage{setspace}
\usepackage{xcolor}
\usepackage{makecell}
\usepackage{ragged2e}

\usepackage{mathtools}

\usepackage[export]{adjustbox}

\newlength{\FigureWidth}
\usepackage{xcolor}
\usepackage{makecell}
\newcommand\TopRule{\Xhline{0.08em}}

\newcommand\MidRule{\Xhline{0.03em}}
\newcommand\BotRule{\Xhline{0.08em}}

\newcommand{\pbu}{{\bar{\psi}}}
\newcommand{\pu}{{\psi}}
\newcommand{\pbd}{{\bar{\chi}}}
\newcommand{\pd}{{\chi}}

\newcommand{\Luv}{{\ell_{\rm UV}}}
\newcommand{\Lir}{{\ell_{\rm IR}}}

\usepackage{relsize}

\usepackage[acronyms, nohypertypes={acronym}, nopostdot, style=super, nonumberlist, toc]{glossaries}
\setacronymstyle{long-sc-short}
\glsenableentrycount
\makeglossaries  

\newcommand\ac[1]{\gls{#1}}

\newcommand\acp[1]{\glspl{#1}}

\newacronym{WF}{wf}{Wilson-Fisher}
\newacronym{QFT}{qft}{Quantum Field Theory}
\newacronym{QCD}{qcd}{Quantum Chromodynamics}
\newacronym{AF}{af}{asymptotically free}
\newacronym{BKT}{bkt}{Berezenski-Kosterlitz-Thouless}

\newacronym{RG}{rg}{renormalization group}

\newacronym{SSF}{ssf}{step-scaling function}

\newacronym{IR}{ir}{infrared}
\newacronym{UV}{uv}{ultraviolet}

\newacronym{QIS}{qis}{Quantum Information Science}
\newacronym{PPT}{ppt}{positive-semidefinite partial transpose}
\newacronym{NPT}{npt}{negative partial transpose}

\definecolor{dkgreen}{rgb}{0.2,0.7,0.4}
\definecolor{dkblue}{rgb}{0.2,0.2,0.7}
\definecolor{dkred}{rgb}{0.8,0,0}
\definecolor{dkgreen}{rgb}{0.2,0.8,0.4}

\begin{document}


\title{Asymptotic Freedom at the Berezinskii-Kosterlitz-Thouless Transition without Fine-Tuning Using a Qubit Regularization}

\author{Sandip Maiti\,\orcidlink{0000-0002-5248-5316}}
\email{sandip.maiti@saha.ac.in}
\affiliation{Saha Institute of Nuclear Physics, HBNI, 1/AF Bidhannagar, Kolkata 700064, India}
\affiliation{Homi Bhabha National Institute, Training School Complex, Anushaktinagar, Mumbai 400094, India}
\author{Debasish Banerjee\,\orcidlink{0000-0003-0244-4337}}
\email{debasish.banerjee@saha.ac.in}
\affiliation{Saha Institute of Nuclear Physics, HBNI, 1/AF Bidhannagar, Kolkata 700064, India}
\affiliation{Homi Bhabha National Institute, Training School Complex, Anushaktinagar, Mumbai 400094, India}
\author{Shailesh Chandrasekharan\,\orcidlink{0000-0002-3711-4998}}
\email{sch27@duke.edu}
\affiliation{Department of Physics, Box 90305, Duke University, Durham, North Carolina 27708, USA}
\author{Marina K.~Marinkovic\,\orcidlink{0000-0002-9883-7866}}
\email{marinama@ethz.ch}
\affiliation{Institut f\"ur Theoretische Physik, Wolfgang-Pauli-Stra{\ss}e 27, ETH Z\"urich, 8093 Z\"urich, Switzerland}

\date{\today}

\begin{abstract}
We propose a two-dimensional hard-core loop-gas model as a way to regularize the asymptotically free
massive continuum quantum field theory that emerges at the Berezinskii-Kosterlitz-Thouless transition.
Without fine-tuning, our model can reproduce the universal step-scaling function of the classical lattice $XY$
model in the massive phase as we approach the phase transition. This is achieved by lowering the fugacity
of Fock-vacuum sites in the loop-gas configuration space to zero in the thermodynamic limit. Some of the
universal quantities at the Berezinskii-Kosterlitz-Thouless transition show smaller finite size effects in our
model as compared to the traditional $XY$ model. Our model is a prime example of qubit regularization of an
asymptotically free massive quantum field theory in Euclidean space-time and helps understand how
asymptotic freedom can arise as a relevant perturbation at a decoupled fixed point without fine-tuning.
\end{abstract}

\maketitle

The success of the Standard Model of particle physics shows that at a fundamental level, nature is well described by a continuum QFT.
Understanding \acp{QFT} non-perturbatively continues to be an exciting area of research, since defining them in a mathematically unambiguous way can be challenging. Most definitions require some form of short-distance (\ac{UV}) regularization, which ultimately needs to be removed. Wilson has argued that continuum \acp{QFT} arise near fixed points of renormalization group flows \cite{Wilson:1983xri}. This has led to the concept of universality, which says that different regularization schemes can lead to the same \ac{QFT}. Following Wilson, traditional continuum quantum field theories are usually regulated non-perturbatively on a space-time lattice by replacing the continuum quantum fields by lattice quantum fields and constructing a lattice Hamiltonian with a quantum critical point where the long distance lattice physics can be argued to be the desired continuum \ac{QFT}. However, universality suggests that there is a lot of freedom in choosing the microscopic lattice model to study a particular \ac{QFT} of interest.

Motivated by this freedom and to study continuum quantum field theories in real time using a quantum computer, the idea of qubit regularization has gained popularity recently \cite{Klco:2018zqz, Alexandru:2019ozf, Singh:2019uwd, Banuls:2019bmf, Zache:2021ggw, Ciavarella:2021nmj, Mariani:2023eix, Zache:2023dko, Hayata:2023puo}. Unlike traditional lattice regularization, qubit regularization explores lattice models with a strictly finite local Hilbert space to reproduce the continuum \ac{QFT} of interest. Euclidean qubit regularization can be viewed as constructing a Euclidean lattice field theory with a discrete and finite local configuration space, that reproduces the continuum Euclidean \ac{QFT} of interest at a critical point. If the target continuum theory is relativistic, it would be natural to explore Euclidean qubit regularized models that are also symmetric under space-time rotations. However, this is not necessary, since such symmetries can emerge at the appropriate critical point. Lattice models with a finite dimensional Hilbert space that can reproduce continuum \acp{QFT} of interest were introduced several years ago through the D-theory formalism  \cite{Wiese:1998nh, Brower:2003vy} and has been proposed for quantum simulations \cite{Banerjee:2012xg, Wiese:2021djl}. In contrast to qubit regularization, the D-theory approach allows the local Hilbert space to grow through an additional dimension when necessary. In this sense, qubit regularization can be viewed as the D-theory approach for those \acp{QFT} where a strictly finite Hilbert space is sufficient to reproduce the desired QFT.

Examples of using qubit regularization to reproduce continuum \acp{QFT} in the \ac{IR} are well known. Quantum spin models with a finite local Hilbert space are known to reproduce the physics of classical spin models with an infinite local Hilbert space near Wilson-Fisher fixed points \cite{sachdev_2011}. They can also reproduce \acp{QFT} with topological terms like the Wess-Zumino-Witten theories \cite{Affleck:1987ch}. Gauge fields have been proposed to emerge dynamically at some quantum critical points of simple quantum spin systems \cite{PhysRevB.70.144407}. From the perspective of Euclidean qubit regularization, recently it was shown that Wilson-Fisher fixed points with $O(N)$ symmetries can be recovered using simple qubit regularized space-time loop models with $N+1$ degrees of freedom per lattice site \cite{Banerjee:2019jpw, Singh:2019jog}. Similar loop models have also been shown to produce other interesting critical behavior \cite{Nahum:2011zd,Nahum:2013xsa,Nahum:2015jya}. Loop models are extensions of dimer models, which are also known to describe interesting critical phenomena in the \ac{IR} \cite{PhysRevE.74.041124,kundu2023flux}. All this evidence shows that Euclidean qubit regularization is a natural way to recover continuum \acp{QFT} that emerge via \ac{IR} fixed points of lattice models.

\begin{figure}[htb]
\includegraphics[width=0.48\textwidth]{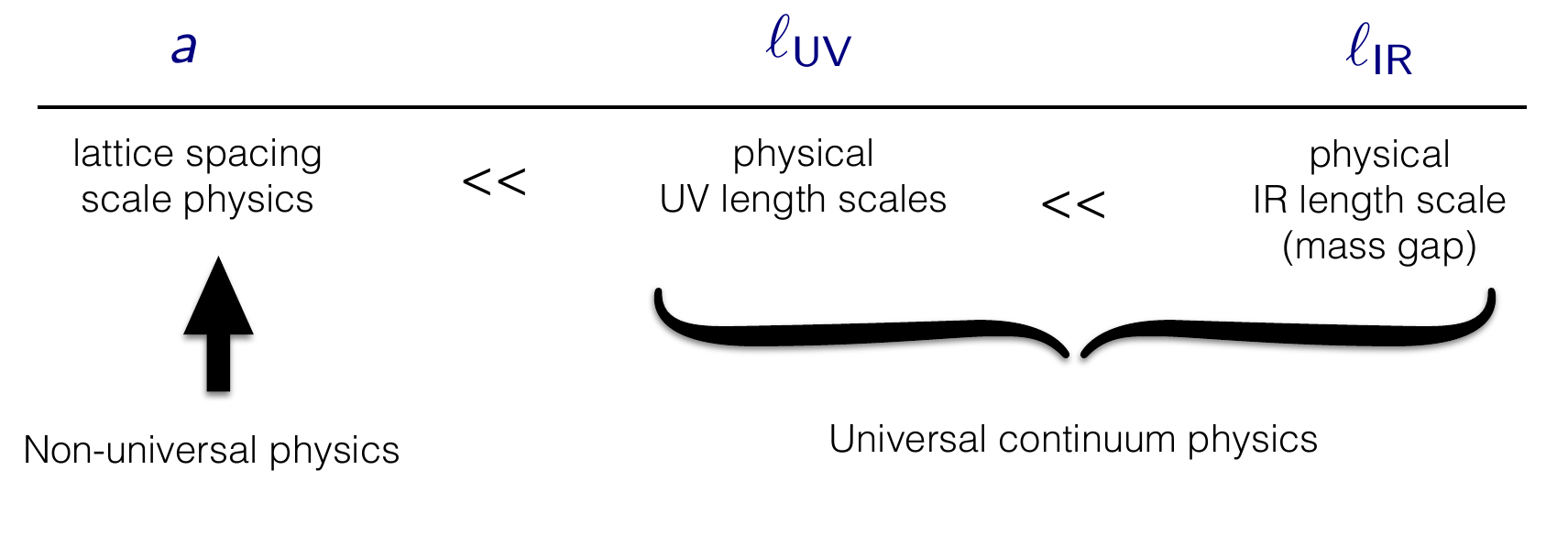}
\caption{\label{fig:afscales}Length scales in a lattice field theory that reproduces asymptotically free quantum field theories.}
\end{figure}

A non-trivial question is whether we can also recover the physics of ultraviolet fixed points (UV-FPs), using qubit regularization. In particular, can we recover massive continuum \acp{QFT} which are free in the UV but contain a marginally relevant coupling? Examples of such \ac{AF} theories include two-dimensional spin models and four dimensional non-Abelian gauge theories. In the D-theory approach, there is strong evidence that the physics at the \ac{UV} scale can indeed be recovered exponentially quickly as one increases the extent of the additional dimension \cite{Bietenholz:2003wa,Beard:2004jr,Laflamme:2015wma,Caspar:2022llo,Zhou:2021qpm}. Can the Gaussian nature of the \ac{UV} theory emerge from just a few discrete and finite local lattice degrees of freedom, while the same theory then goes on to reproduce the massive physics in the \ac{IR}? For this we will need a special type of quantum criticality where three length scales, as sketched in \cref{fig:afscales}, emerge. There is a short lattice length scale $a$, where the non-universal physics depends on the details of the qubit regularization, followed by an intermediate length scale $\Luv \gg a$, where the continuum \ac{UV} physics sets in and the required Gaussian theory emerges. Finally, at long length scales $\Lir \gg \Luv$, the non-perturbative massive continuum quantum field theory emerges due to the presence of a marginally relevant coupling in the \ac{UV} theory. The qubit regularized theory thus reproduces the universal continuum \ac{QFT} in the whole region $\ell_{\rm UV}$ to $\ell_{\rm IR}$. The special quantum critical point must be such that $\ell_{\rm UV}/a \rightarrow \infty$. 

\begin{figure}[htb]
\includegraphics[width=0.4\textwidth]{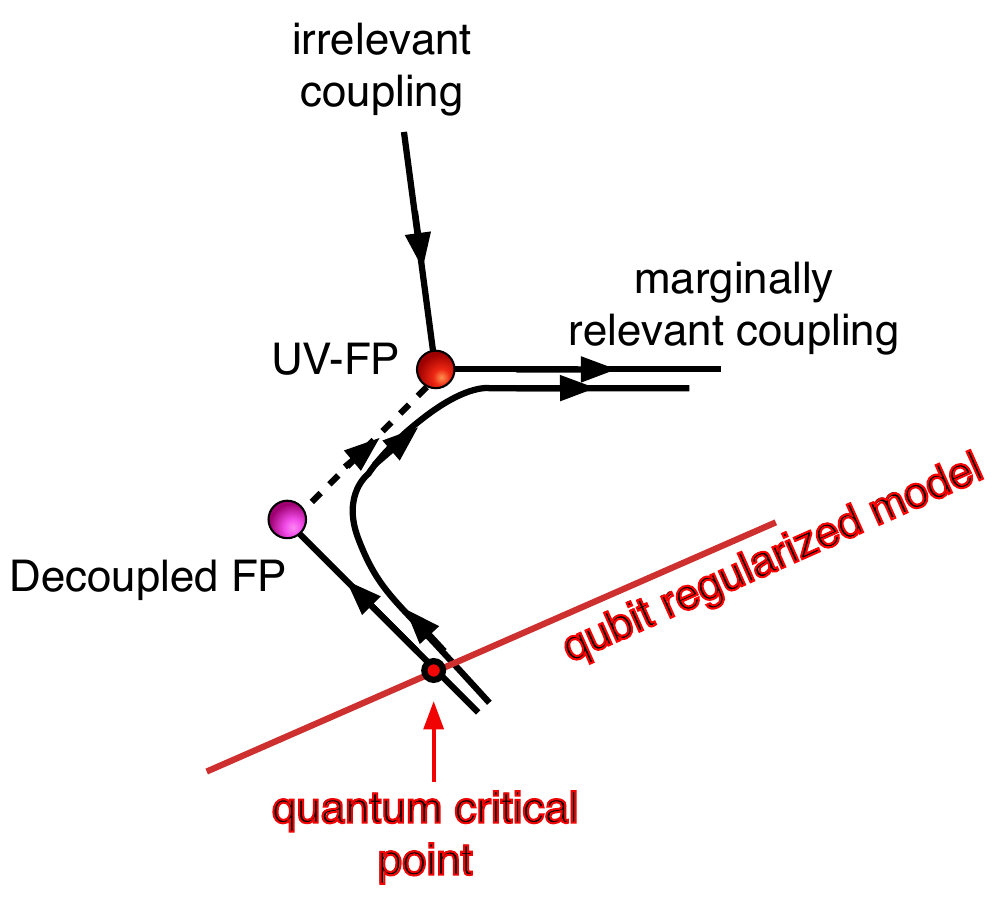}
\caption{An illustration of the RG flow of a generic qubit regularized model that reproduces the physics of the asymptotically free \acp{QFT}. At the decoupled quantum critical point, qubit models describe the physics of a critical system containing two decoupled theories. However, when a small non-zero coupling is introduced between the theories, the long distance physics flows towards the desired universal physics of the \ac{UV}-fixed point theory.
\label{fig:RG-flow}}
\end{figure}

Recently, a quantum critical point with these features was discovered in an attempt to find a qubit regularization of the asymptotically free massive non-linear O(3) sigma model in two space-time dimensions in the Hamiltonian formulation \cite{Bhattacharya:2020gpm}. Using finite size scaling techniques, it was shown that the qubit regularized model recovers all the three scales. In this paper, we report the discovery of yet another example of a quantum critical point with similar features. In the current case, it is a Euclidean qubit regularization of the asymptotically free massive continuum quantum field theory that arises as one approaches the \ac{BKT} transition from the massive phase \cite{Berezinsky:1970fr,Kosterlitz:1973xp}. In both these examples, the qubit regularized model is constructed using two decoupled theories and the \ac{AF}-\ac{QFT} emerges as a relevant perturbation at a decoupled quantum critical point. The coupling between the theories plays the role of the perturbation that creates the three scales. A generic RG flow diagram of such qubit regularized theories is illustrated in \cref{fig:RG-flow}. An interesting feature of this discovery is that there is no need for fine-tuning to observe some of the universal features of the \ac{BKT} transition that have been unattainable in practice with other traditional regularizations \cite{Kenna:1996bs}.

\begin{figure*}[t]
\includegraphics[width=0.48\textwidth]{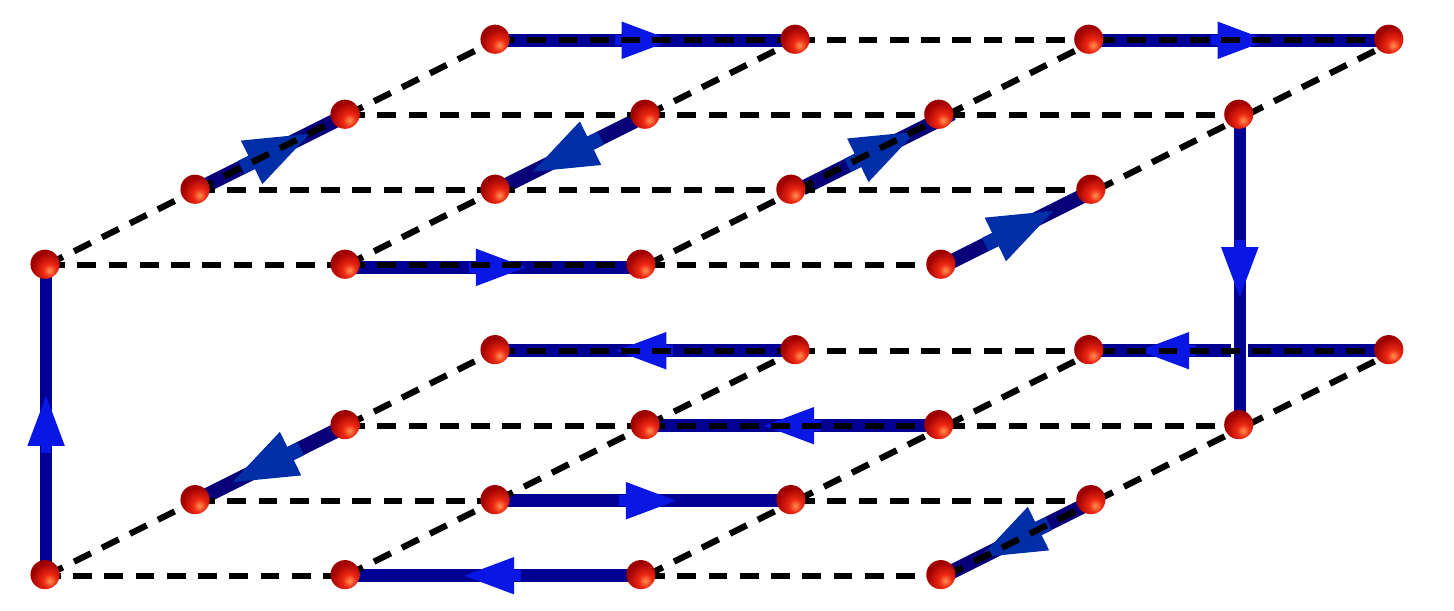}
\hskip1in
\includegraphics[width=0.2\textwidth]{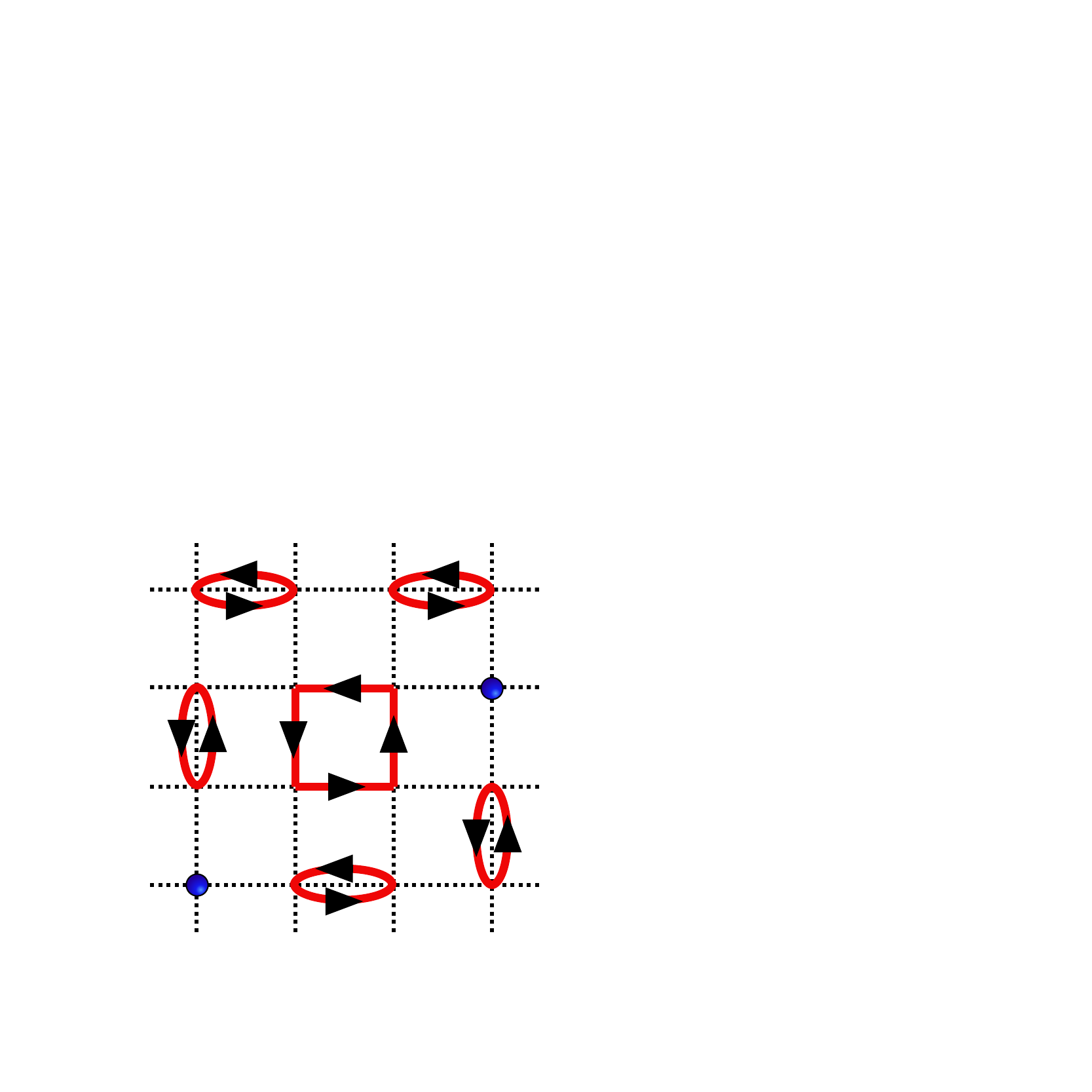}
\caption{\label{fig:loopmodel} The left figure shows an illustration of a dimer configuration that contributes to the partition function of the model arising from \cref{eq:dimermodel}. Interlayer dimers (or instantons) have weight $\lambda$ while the intralayer dimers have weight one. By giving the dimers an orientation as illustrated, each dimer configuration can also be viewed as a configuration of self-avoiding oriented loops. The configuration on the right is such a mapping of the configuration on the left. The instantons are mapped into Fock-vacuum sites, shown as blue circles. The dimer model shows that the loop model is critical when $\lambda=0$.}
\end{figure*}

The \ac{BKT} transition is one of the most widely studied classical phase transitions, since it plays an important role in understanding the finite temperature superfluid phase transition of two-dimensional systems \cite{RevModPhys.85.1191}. One simple lattice model that captures the universal behavior of the physics close to the phase transition is the 
classical two-dimensional XY model on a square lattice given by the classical action,
\begin{align}
S = -\beta \sum_{\langle ij\rangle} \cos(\theta_i-\theta_j),
\label{eq:xymodel}
\end{align}
where the lattice field $0\leq \theta_i < 2\pi$ is an angle associated to every space-time lattice site $i$ and $\langle ij\rangle$ refers to the nearest neighbor bonds with sites $i$ and $j$. We refer to this as the $\mathrm{bXY}$ model. The lattice field naturally lives in an infinite dimensional Hilbert space of the corresponding one dimensional quantum model. Using high precision Monte Carlo calculations, the \ac{BKT} transition has been determined to occur at the fine-tuned coupling of $\beta_c \approx 1.1199(1)$ \cite{Hasenbusch:2005xm,Hasenbusch:2006vd}. 
The BKT phase transition has also been studied using the Villain model \cite{PhysRevB.16.1217}, which is better suited to analytic computations, as well topological actions \cite{Bietenholz:2012ud}.
While the above approaches to the \ac{BKT} transition require fine-tuning, the massive phase near the transition can be reached without fine-tuning through fermionic models \cite{ZJ2002}. It was recently shown how the two-flavor Schwinger model at $\theta=\pi$ reproduces the exponentially small mass-gap expected near the \ac{BKT} transition~\cite{dempsey2023phase}. The model we consider in this work is similar, but without explicit gauge fields and constructed via hard-core bosons.

As one approaches the \ac{BKT} transition from the massive phase, the long distance physics of the \cref{eq:xymodel} is known to be captured by the sine-Gordon model whose Euclidean action is given by\cite{ZJ2002},
\begin{align}
S = \int dx d\tau\ \left[ \frac{1}{2t} (\partial_\mu \theta_1)^2 + \frac{t}{8\pi^2} (\partial_\mu \theta_2)^2 - 
\frac{A t}{4\pi^2}\cos\theta_2 \right]
\label{eq:BKTqft}
\end{align}
where $t \geq \pi/2$. The field $\theta_1(x,\tau)$ captures the spin-wave physics while the vortex dynamics is captured by the field $\theta_2(x,\tau)$. The BKT transition in this field theory language occurs at $t = \pi/2$ where the $\cos\theta_2$ term becomes marginal as one approaches the critical point and the physics is governed by a free Gaussian theory. In this sense, at length scales much larger than the lattice spacing, the  physics of the lattice XY model is the same as an asymptotically free massive Euclidean continuum \ac{QFT}, when
$\beta$ is tuned to $\beta_c$ from smaller values.

Qubit regularizations of the classical XY-model have been explored recently using various quantum spin formulations \cite{Zhang:2021dnz}. Lattice models based on the spin-1 Hilbert space are known to contain rich phase diagrams \cite{PhysRevB.67.104401}, and quantum field theories that arise at some of the critical points can be different from those that arise at the \ac{BKT} transition. Also, the presence of a marginally relevant operator at the BKT transition can make the analysis difficult, especially if the location of the critical point is not known. In these cases, it becomes a fitting parameter in the analysis, increasing the difficulty. Since in our model the location of the critical point is known, our model can be analyzed more easily.

\begin{figure}[!t]
\includegraphics[width=0.48\textwidth]{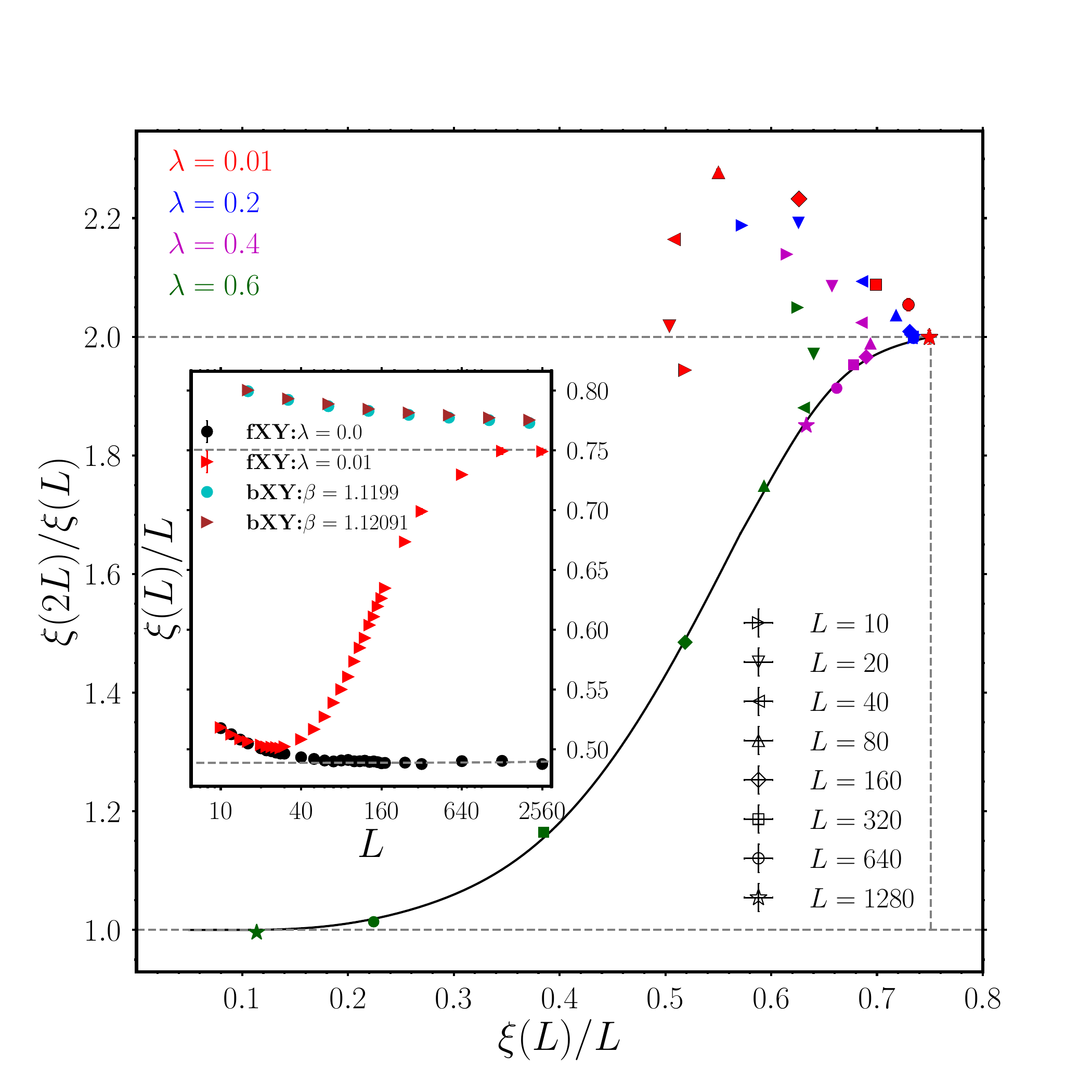}
\caption{The figure shows the universal \ac{SSF} (i.e., $\xi(2L)/\xi(L)$ vs. $\xi(L)/L$) obtained from the XY-model \cref{eq:xymodel} (solid line) \cite{SupMat} and compares it with data from the model \cref{eq:dimermodel} at $\lambda=0.01$ (red), $0.2$(blue), $0.4$(purple) and $0.6$ (green), for various lattice sizes shown with different symbols. 
{For small values of $L$, our data deviate from the solid line. We define $\Luv$ as the minimum value of $L$ when the data begin to fall on the solid line. From the figure we estimate $\Luv \approx 80$ for $\lambda=0.6$ and $\Luv \approx 160$ for $\lambda=0.4$.} For very small $\lambda$ we expect the $\xi(L)/L$ to approach the universal UV prediction of $\xi(L)/L=0.7506912...$ (see \cite{Hasenbusch:2005xm}), when $L \sim \Luv$ before beginning to follow the solid line. We see this at $\lambda=0.2$ and $0.01$. 
{Since at these couplings $\Luv > 1280$,} we predict that the data at these couplings will also eventually follow the solid line, but only for $L \gg \Luv$ which we cannot access. To show this feature, in the inset we plot $\xi(L)/L$ as a function of $L$ at $\lambda=0.01$. Note that the data approaches $\xi(L)/L=0.7506912...$ when $L\sim \Luv$ as expected. Based on our prediction above, this is only a plateau and that for $L \gg \Luv$ (which we cannot access) $\xi(L)/L$ will eventually approach zero. The inset also shows that the large $L$ behavior of $\lambda=0$ is very different and stabilizes at $\xi(L)/L=0.4889(6)$. In the inset we also show the data from \cite{Hasenbusch:2005xm} in the traditional $XY$ model (\cref{eq:xymodel}) at two values of $\beta$ close to the transition. This data is still far from the universal value due to logarithmic corrections as explained in \cite{Hasenbusch:2005xm}.}
\label{fig:chi2L}
\end{figure}

\begin{figure}[ht]
\includegraphics[width=0.48\textwidth]{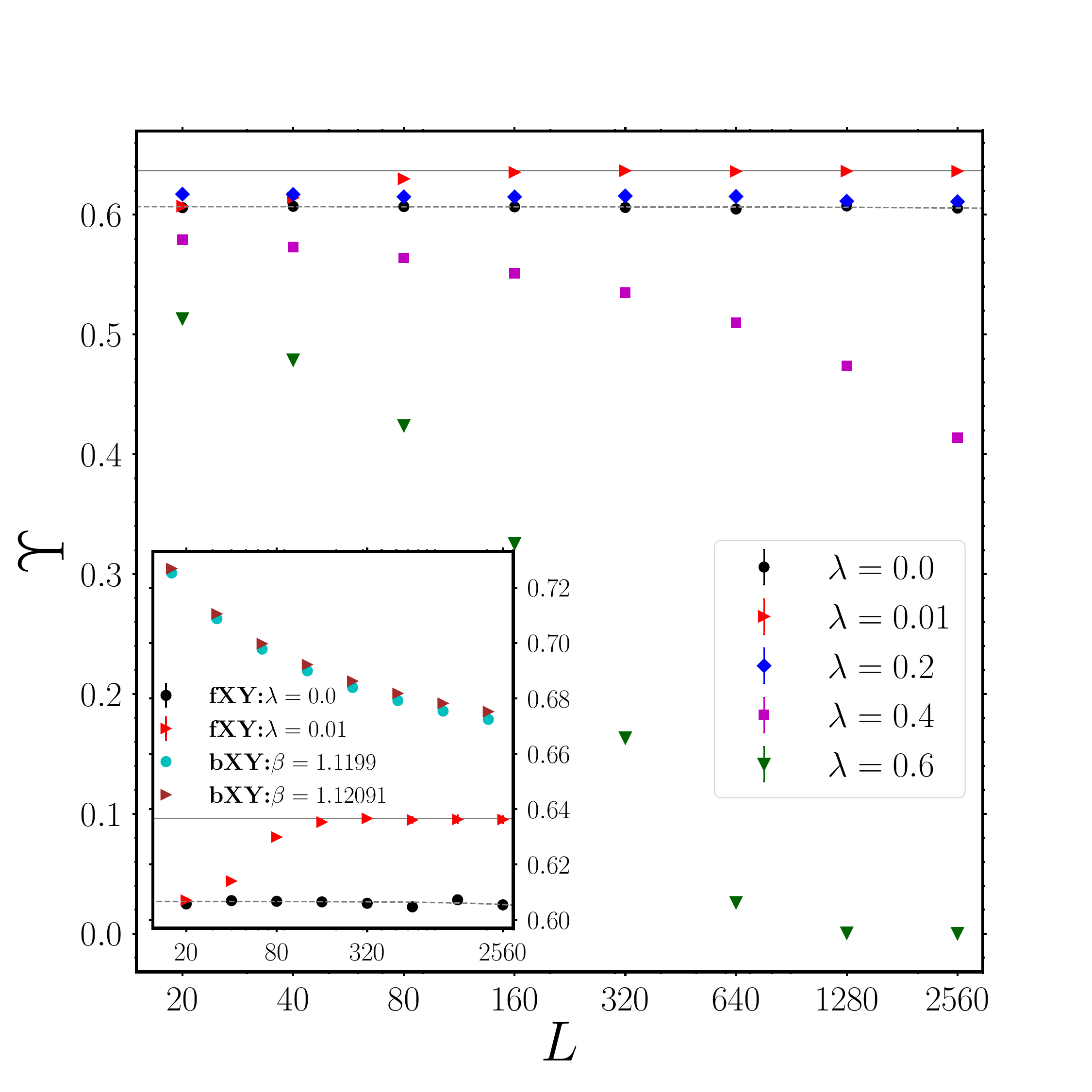}
\caption{The figure shows the helicity modulus $\Upsilon$ as a function of $L$ for $\lambda=0.6$, $0.4$, $0.2$,$0.01$ and $0.0$. We expect $\Upsilon \rightarrow 0$ for $L \gg \Luv$ when $\lambda \neq 0$. This is clearly seen at $\lambda = 0.6$. At $\lambda=0.4$, since $\Luv$ is larger, we only see the initial part of the decrease towards zero. At $\lambda=0.2$, $\Luv$ is even larger, and so we only observe the flat part expected in the \ac{UV}. At the \ac{UV} fixed point we expect $\Upsilon \approx 2/\pi$. When $\lambda = 0.01$ we do observe the data converging well to this universal value (top solid line). Since $\lambda=0.01$ is not really the critical point, this line too will eventually turn around at very large values of $L$ and go to zero. On the other hand when $\lambda=0$ we find that $\Upsilon \approx 0.606$ in the large $L$ limit \cite{SupMat}. We demonstrate the difference between $\lambda=0$ and $\lambda=0.01$ in the inset, where we also show data from \cite{Hasenbusch:2005xm} for the helicity modulus in the traditional XY model \cref{eq:xymodel} at two values of $\beta$ close to the transition.
Note that the values from the traditional model are far from $2/\pi$, a well-known difficulty due to large logarithmic corrections. In contrast, our qubit model is able to recover the \ac{UV} physics more easily.}
\label{fig:qw}
\end{figure}

The model we consider in this work can be compactly written in terms of the Grassmann integral
\begin{align}
Z \ = \int\ &[d\pbu d\pu] \ [d\pbd d\pd] \ \exp\Big(\lambda \ \sum_i \ \pbu_i \pu_i \pbd_i\pd_i\Big)
\nonumber \\
& \times\ \exp\Big( \sum_{\langle ij\rangle} \big( \pbu_i \pu_i \pbu_j\pu_j + \pbd_i \pd_i \pbd_j \pd_j\big)\Big)
\label{eq:dimermodel}
\end{align}
where on each site $i$ of the square lattice we define four Grassmann variables $\pbu_i$, $\pu_i$, $\pbd_i$ and $\pd_i$ and assume periodic lattices with $L$ sites in each direction. We refer to \cref{eq:dimermodel} as the fXY model. Using the fermion bag approach \cite{PhysRevD.82.025007}, we can integrate the Grassmann variables and write the partition function as a sum over dimer configurations whose weight is given by $\lambda^{N_I}$ where $N_I$ is the number of instantons (or Fock-vacuum sites). An illustration of such a configuration is given in \cref{fig:loopmodel}. The inter-layer dimers resemble t'Hooft vertices in the fermionic theory \cite{Ayyar:2014eua, Ayyar:2015lrd, Maiti:2021wqz}.
Thus, $\lambda$ plays the role of the fugacity of instantons. It is easy to verify that the action of our model is invariant under $\pbu_j \pu_j \rightarrow e^{i\sigma_j\theta} \pbu_j \pu_j$ and $\pbd_j \pd_j \rightarrow e^{-i\sigma_j\theta} \pbd_j \pd_j$ where $\sigma_j = \pm$ tracks the parity of the site $j$. The critical behaviour of this U(1) symmetry is connected to the \ac{BKT} transition.

The configurations in \cref{fig:loopmodel} can also be viewed as configurations of oriented self-avoiding loops on a square lattice with Fock-vacuum sites if the dimers are given orientation as explained in the caption of the figure. The model we consider in this work is a variant of the qubit regularized XY model introduced in Euclidean space recently \cite{Singh:2019uwd}. The loop model can be viewed as a certain limiting case of the classical lattice XY-model \cref{eq:xymodel} written in the world-line representation \cite{Banerjee:2010kc}, where the bosons are assumed to be hard-core. The main difference between our model in this work and the one introduced in \cite{Singh:2019uwd} is that closed loops on a single bond are now allowed. Such loops seemed unnatural in the Hamiltonian framework that motivated the previous study, but seem to have profoundly different features in two dimensions because it is possible to view the loop configurations as a configuration of closed packed oriented dimers and argue for a critical point in our model at $\lambda=0$ and a massive phase for $\lambda>0$. The previous model does not have this property \cite{Singh-Comm}. 

Using worm algorithms (see \cite{Adams:2003cca}) we study our model for various values of $L$ and $\lambda$.
At $\lambda = 0$, one gets two decoupled layers of closed packed dimer models, which is known to be critical \cite{Kast1963,PhysRev.124.1664,PhysRev.132.1411,ALLEGRA2015685}. The effect of $\lambda \neq 0$ was studied several years ago, and it was recognized that there is a massive phase for sufficiently large values of $\lambda$ \cite{PhysRevD.68.091502,Wilkins2020}. However, the scaling of quantities as $\lambda \rightarrow 0$ was not carefully explored. 
{Recently, the subject was reconsidered
\cite{PhysRevE.103.042136}, and the emergence of a long crossover phenomenon was discovered for small $\lambda$ as a function of $L$. However, the universal properties of this crossover being related to the UV physics at the BKT transition was not appreciated.}
In this paper, we demonstrate that the observed crossover phenomena captures the asymptotic freedom of \cref{eq:BKTqft}. We do this by comparing the universal behavior of \cref{eq:dimermodel} with the traditional XY model \cref{eq:xymodel} near the massive phase of the \ac{BKT} transition \cite{Hasenbusch:2005xm,Balog:2000ra,Balog:2001wv}.

To compare universal behaviors of \cref{eq:xymodel} and \cref{eq:dimermodel} we compute the second moment finite size correlation length $\xi(L)$ defined as $\xi(L) \ =\ \sqrt{(\chi/F)-1}/(2\sin(\pi/L))$ (see \cite{Caracciolo:1994ud}), where $\chi = G(0)$ and $F = G(2\pi/L)$ are defined through the two point correlation function
\begin{align}
G(p) = \sum_j e^{i p x} \langle {\cal O}^+_{(x,\tau)} {\cal O}^-_{(0,0)}\rangle.   
\end{align}
In the above relation $j$ is the space-time lattice site with coordinates \textcolor{red}{$(x,\tau)$} and ${\cal O}^+_j, {\cal O}^-_j$ are appropriate lattice fields in the two models. In the $XY$ model ${\cal O}^+_j = e^{i\theta_j}, {\cal O}^-_j = e^{-i\theta_j}$, while in the dimer model ${\cal O}^+_j = {\cal O}^-_j = \pbu_j \pu_j$. We demonstrate that the step-scaling function (SSF) (i.e., the dependence of $\xi(2L)/\xi(L)$ on $\xi(L)/L$) of the two lattice models show excellent agreement with each other in the scaling regime $\ell_{UV} \gg a$, in \cref{fig:chi2L}.

Another interesting universal result at the \ac{BKT} transition is the value of the helicity modulus, which can be defined using the relation, $\Upsilon\ =\ \langle Q_w^2\rangle$ where $Q_w$ is the spatial winding number of bosonic worldlines. In the XY model \cref{eq:xymodel}, it is usually defined using a susceptibility of a twist parameter in the boundary conditions \cite{Hasenbusch:2005xm}. In our model, we can easily compute the winding charge $Q_w$ in each loop configuration illustrated in \cref{fig:loopmodel}. The universal result in the massive phase as we approach the BKT transition is that $\Upsilon \approx 2/\pi$ in the \ac{UV} up to exponentially small corrections \cite{Hasenbusch:2005xm}, although in the \ac{IR} $\Upsilon = 0$. While it is difficult to obtain the \ac{UV} value in lattice calculations using the traditional model \cref{eq:xymodel}, in our model, we can see it emerge nicely at $\lambda=0.01$. We demonstrate this in \cref{fig:qw}. Again, as expected, the value of $\Upsilon$ when $\lambda=0$ is very different, since it is a theory of free bosons but at a different coupling. Using the different value of the coupling gives $\Upsilon\ \approx 0.606$ \cite{SupMat}. Our results provide strong evidence that the \ac{AF}-\ac{QFT} at the BKT transition emerges from our dimer model when we take the limit $L\rightarrow \infty$ followed by $\lambda\rightarrow 0$. The opposite limit leads to the critical theory of the decoupled dimer model.

{\bf Acknowledgments:} We are grateful to F.~Alet, J.~Pinto Barros, S. Bhattacharjee, T. Bhattacharya, K. Damle, I. Klebanov, H. Liu, S. Pujari, A. Sen, H. Singh and U.-J. Wiese for inspiring discussions. We acknowledge use of the computing clusters at SINP, and the access to Piz Daint at the Swiss National Supercomputing Centre, Switzerland under the ETHZ’s share with the project IDs go24 and eth8. Support from the Google Research Scholar Award in Quantum Computing and the Quantum Center at ETH Zurich is gratefully acknowledged. S.C's contribution to this work is based on work supported by the U.S. Department of Energy, Office of Science --- High Energy Physics Contract KA2401032 (Triad National Security, LLC Contract Grant No. 89233218CNA000001) to Los Alamos National Laboratory. S.C is supported by a Duke subcontract based on this grant. S.C's work is also supported in part by the U.S. Department of Energy, Office of Science, Nuclear Physics program under Award No. DE-FG02-05ER41368.

\bibliography{ref}

\clearpage

\appendix

\begin{center}
{\bf \Large Supplementary Material}    
\end{center}

\section{I. Universal values of $\Upsilon$ for $\lambda = 0$ and $\lambda \neq 0$}

In this section we explain the two different values of the helicity modulus $\Upsilon$ for our model when $\lambda=0$ and $\lambda \rightarrow 0$. When $\lambda=0$ our model maps into two identical but decoupled layers of close-packed classical dimer models. As has already been explained in the literature (see for example ~\cite{PhysRevE.74.041124,ALLEGRA2015685}), each layer can be mapped to the theory of a free compact scalar field with the action 
\begin{equation}
S = \frac{1}{2 t} \int d^2 x (\partial_\mu \theta(x))^2.
\label{eq:freeboson}
\end{equation}
with $t=4\pi$. One can compute $\Upsilon$ starting with \cref{eq:freeboson}, by noting that the scalar fields have winding number configurations labeled by $n_x$:
\begin{equation}
\theta(x) \ =\ \frac{2 \pi x n_x}{L_x} + \varphi(x),
\end{equation}
where $\varphi(x)$ is a smooth fluctuation that is independent of winding number $n_x$. The value of the action in each winding sector in a finite space-time volume is then given by
\begin{equation}
S(n_x) = \frac{2\pi^2 n_x^2}{t} \frac{L_y}{L_x} + S_0,
\end{equation}
where $S_0$ is the action from the usual fluctuations in the zero winding number sector. Using $L_x = L_y$, we can compute $\Upsilon$ using its connection to the average of the square of the winding numbers,
\begin{align}
\Upsilon\ =\ \braket{ (Q_x)^2 }  = \frac{ \sum_{n_x} n_x^2 \cdot {\rm e}^{- \frac{2 \pi^2 n_x^2}{t}} }{ \sum_{n_x} {\rm e}^{-\frac{2\pi^2 n_x^2}{t}}}
\label{eq:Upsilon1}
\end{align}
Numerically evaluating this expression for $t=4\pi$ we obtain $\Upsilon \ = 0.303426...$ for a each layer of our dimer model. Our value of $ 0.606852...$ is due to the presence of two decoupled layers.

In contrast, in the limit $\lambda \rightarrow 0$, we need to consider the physics at the BKT transition and so we begin with the  action 
\begin{align}
S = \int d^2x\ \left[ \frac{1}{2\tilde{t}} (\partial_\mu \theta_1)^2 + \frac{\tilde{t}}{8\pi^2} (\partial_\mu \theta_2)^2 - 
\frac{A \tilde{t}}{4\pi^2}\cos\theta_2 \right]
\end{align}
and focus at $\tilde{t}=\pi/2$. At this coupling the last term is irrelevant and $\Upsilon$ gets dominant contribution from the $\theta_2$ field. In this we can still use \cref{eq:Upsilon1} but need to substitute $t = 4\pi^2/\tilde{t} = 8\pi$. Substituting we get $\Upsilon = 0.636508...$ which is approximately $2/\pi$.

\section{II. Worm Algorithm}
In this section, we discuss the worm algorithm we use to simulate the model with the partition function,
\begin{align}
Z \ = \int\ &[d\pbu d\pu] \ [d\pbd d\pd] \ \exp\Big(\lambda \ \sum_i \ \pbu_i \pu_i \pbd_i\pd_i\Big)
\nonumber \\
& \times\ \exp\Big( \sum_{\langle ij\rangle} \big( \pbu_i \pu_i \pbu_j\pu_j + \pbd_i \pd_i \pbd_j \pd_j\big)\Big)
\label{eq:2DimerModel}
\end{align}
as introduced in the main paper. These algorithms are well known \cite{Adams:2003cca}, and can be divided into three parts: {\bf Begin}, {\bf Move}, and {\bf End}. 

\begin{enumerate}
\item {\bf Begin}: pick a site at random and denote it as \emph{tail}, and there are the following
  two possibilities: (A) either it has a bond connected to it on the other layer (which we call
  an instanton, or an interlayer dimer), or, (B) it has a bond connected to it on the same layer 
  (which we call a dimer).

  \begin{itemize}
  \item For the case (A), propose to remove the instanton, and put the worm \emph{head} on the 
  same site at the different layer, with a probability $1/\lambda$. If accepted, then begin the 
  worm update, otherwise go to (1).
  
  \item For the case (B), pick the other site to which the dimer is connected as the \emph{head},
  and begin the worm update. 
  \end{itemize}

\begin{table*}[htb]
\renewcommand{\arraystretch}{1.3}
\setlength{\tabcolsep}{1.9pt}
    \centering
\begin{tabular}{l|cc|cc|cc r@{.}l c}
  \hline
$\lambda$ & \multicolumn{2}{c|}{$\rho$} & \multicolumn{2}{c|}{$\chi_1$}  & \multicolumn{2}{c}{$\Upsilon$} \\
   \hline
     &  Exact & Worm &  Exact & Worm  & Exact & Worm\\
  \hline  
0.0  & 0 & 0 & 0.25000 & 0.25004(4) & 0.5 & 0.5(0)\\
0.01 & 0.00001 & 0.00001(0) & 0.25000 & 0.25000(4) & - & - \\
0.2  & 0.00498 & 0.00498(1) & 0.24876 & 0.24877(4) & - & - \\
0.4  & 0.01961 & 0.01963(3) & 0.2451  & 0.24510(4) & - & - \\
0.6  & 0.04306 & 0.04304(5) & 0.23923 & 0.23923(4) & - & - \\
  \hline
\end{tabular}
\caption{ A comparison table between exact and Monte Carlo results on a  $2 \times 2$ lattice.}
 \label{Tab:I}
\end{table*}

\item {\bf Move}: Propose to move the worm head to one of the $(2D+1)$ neighbor sites of \emph{head} 
 with an equal probability, which can either be on the same layer ($2D$ choices), or on the different 
 layer (one choice). Denote the proposed new site as \emph{site0}, and the following possibilities can 
 occur, provided that \emph{site0} is not the \emph{tail}: 

        \begin{itemize}   

         \item \emph{site0} is on the same layer, and has an instanton connected to it. Propose to 
         remove the instanton with a probability $1/\lambda$. If accepted, place the \emph{head} 
         at \emph{site0}, but on the different layer. 
        
        \item \emph{site0} is on the same layer, and has a dimer connected to it (joining \emph{site0}
        and $y$). Move the \emph{head} to the site $y$ with a probability 1, and simultaneously insert 
        a dimer between \emph{head} and \emph{site0}. 
        
        \item \emph{site0} is on the different layer, then propose if an instanton can be created. If
        yes, then move the position of the \emph{head} to $y$ in the other layer, where $y$ is the other
        end of the dimer connecting \emph{site0} and $y$. 
        
        \end{itemize}

\item {\bf End}: If at any stage in the algorithm, the \emph{site0} is the \emph{tail}, then propose to end 
 the worm update. If the \emph{site0} $=$ \emph{tail} is on the same layer, then end the update by putting 
 a dimer between the \emph{head} and \emph{tail} with a probability 1. If, on the other hand, they are
 on different layers, the worm update ends with a probability $\lambda$, leading to the addition of an
 extra instanton. 
\end{enumerate}

\section{III. Exact vs Monte Carlo results on a $2 \times 2$ lattice}
In this work, we compute two independent fermion bilinear susceptibilities defined as
\begin{align}
\chi_1 =  \frac{1}{2V} \sum\limits_{\substack{i,j \\ i\neq j}} \left\langle \bar \psi_i \psi_i \bar \psi_j \psi_j \right\rangle,
\end{align}
\begin{align}
\chi_2 =  \frac{1}{2V} \sum\limits_{\substack{i,j \\ i\neq j}} \left\langle \bar \psi_i \psi_i \bar \chi_j \chi_j \right\rangle,
\end{align}
where $\chi_1$ is an observable that can be defined even on a single layer, while $\chi_2$ is involves both the layers. When the coupling $\lambda = 0$, the two layers are completely decoupled from each other and we get $\chi_2 = 0$. Another quantity we compute is the average density of Fock vacuum sites or inter-layer dimers (which we also view as instantons), defined as
 \begin{align}
 \rho = \frac{1}{V} \sum\limits_i \left\langle \bar \psi_i \psi_i \bar \chi_i \chi_i \right\rangle,
\end{align}
where the expectation value is defined as 
\begin{align}
     \left\langle {\cal O} \right\rangle = \frac{1}{Z} \int [\mathcal{D}\bar \psi \mathcal{D} \psi] \hspace{0.1cm}
     [\mathcal{D}\bar \chi \mathcal{D} \chi]\hspace{0.1cm} {\cal O} \hspace{0.1cm} 
     e^{-S[\bar \psi,\psi, \bar \chi,\chi]}.
\end{align}
Since every site is populated by either a Fock-vacuum site or an intra-layer dimer, the average intra-layer dimer density is not an independent observable. We can always compute it from the Fock vacuum sites (instanton) density $\rho$. 

In order to test out algorithm, we focus on exact results on a $2\times 2$ lattice. The partition function in this simple case is given by
\begin{align}
    Z = 64 + 16 {\lambda}^2 + {\lambda}^4,
\end{align}
while the instanton density and the two independent susceptibilities are given by
\begin{align}
    \rho &= \frac{1}{4Z} (32 {\lambda}^2 + 4 {\lambda}^4), \\
    \chi_1 &= \frac{1}{2Z} (32 + 4 {\lambda}^2), \\
    \chi_2 &= \frac{1}{2Z} (8 {\lambda}).
\end{align}
Note that $\rho$ is zero when $\lambda = 0$ and approaches one for large couplings. Also, as expected $\chi_2=0$ when $\lambda=0$. In \cref{Tab:I} we compare results for three different observables, instanton density ($\rho$), fermion bilinear susceptibility ($\chi_1$), and helicity modulus ($\Upsilon$) on a $2 \times 2$ lattice obtained from an exact calculation against the results obtained using the worm algorithm.

\begin{figure*}
\includegraphics[width=0.46\textwidth]{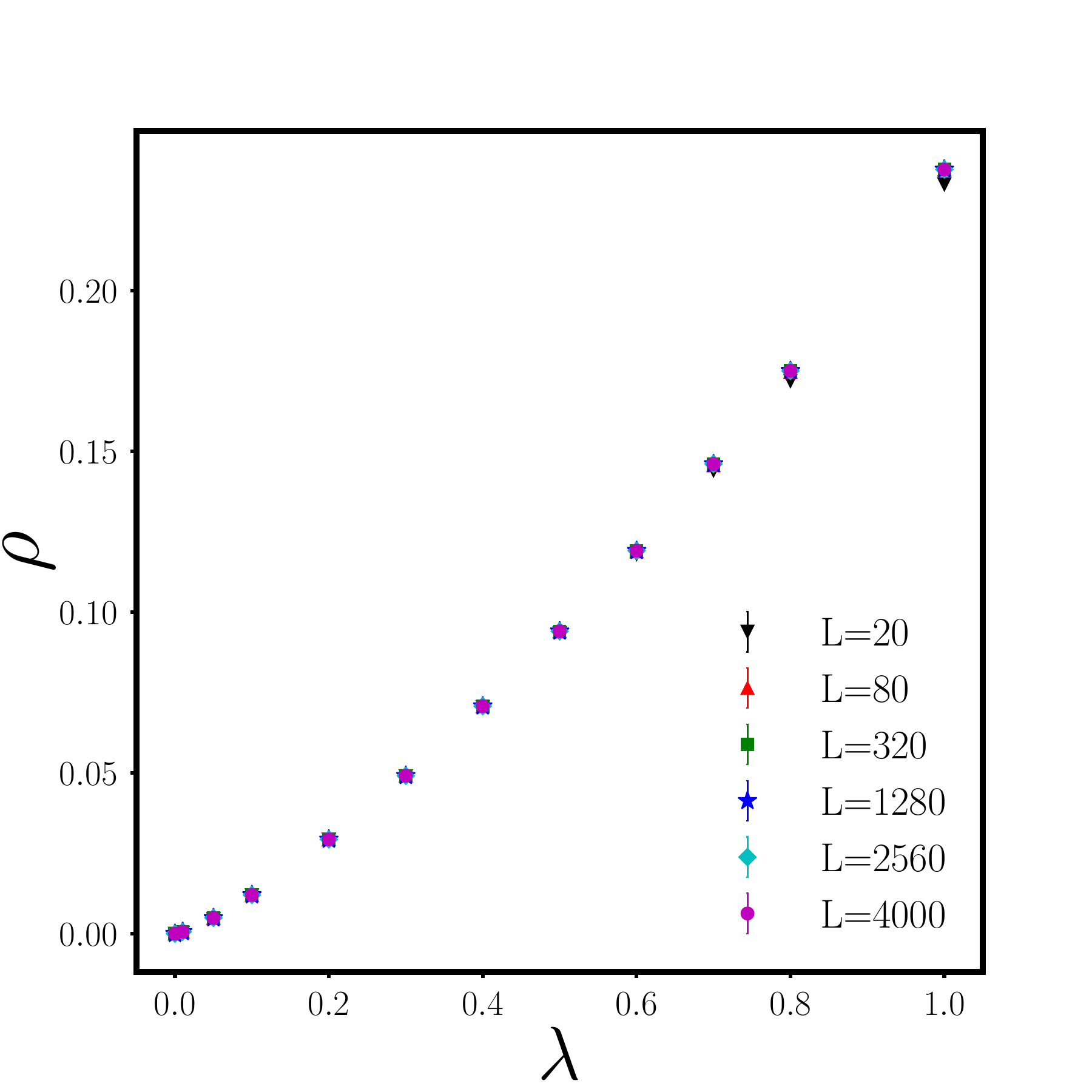}
\includegraphics[width=0.46\textwidth]{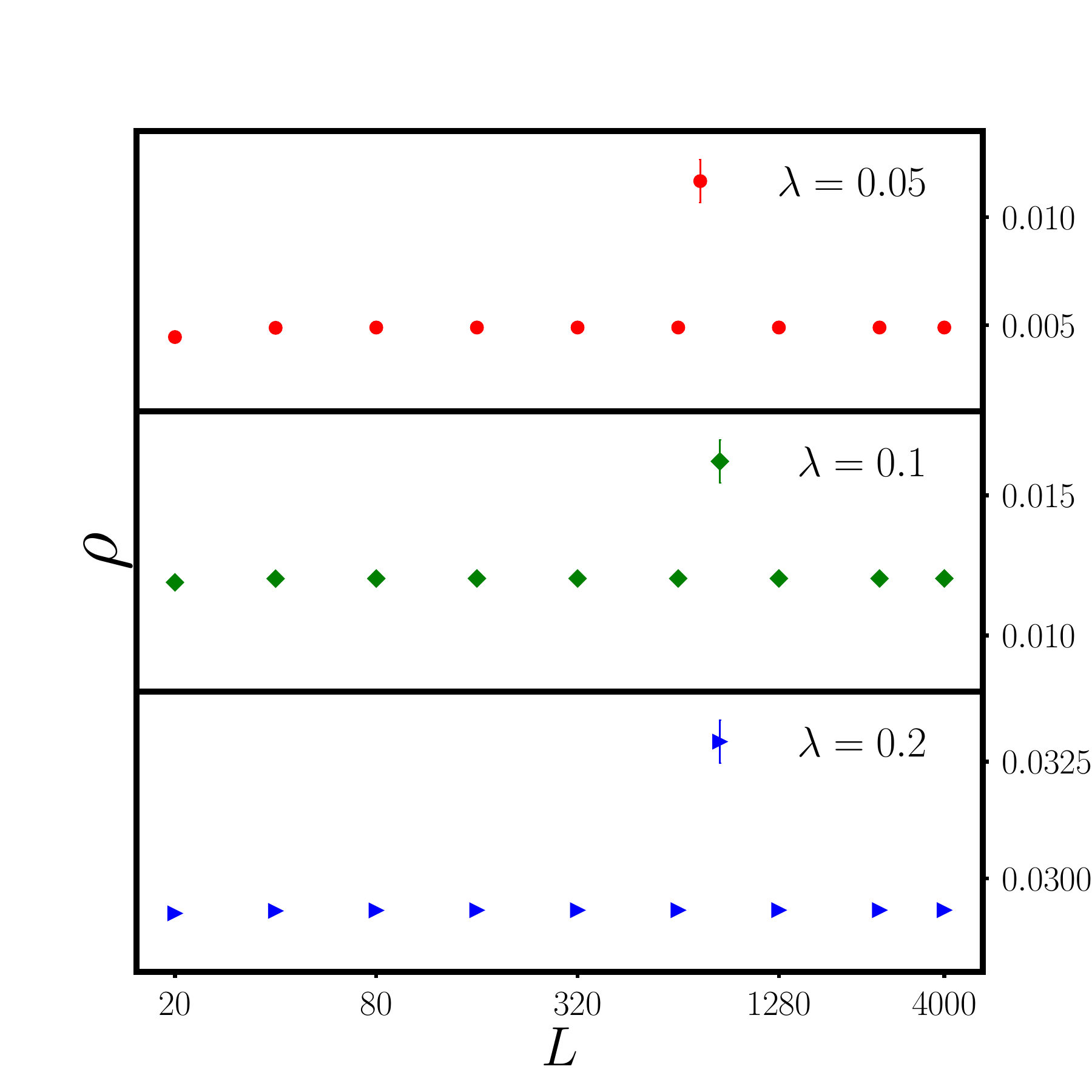}
\caption{On the left we plot the variation of $\rho$, as a function of $\lambda$ for different lattice sizes. On the right, we show the variation in $\rho$ as a function of $L$ for couplings $\lambda=0.05, 0.1$, and $0.2$.  }
\label{fig:rho}
\end{figure*}

Interestingly, when $\lambda \neq 0$ we find that both $\chi_1$ and $\chi_2$ become similar as $L$ increases. The difference also becomes smaller as $\lambda$ increases. We show this behavior in the \cref{Tab:Ia}.
\begin{table}[!htb]
\renewcommand{\arraystretch}{1.3}
\setlength{\tabcolsep}{1.9pt}
    \centering
\begin{tabular}{l|cc|cc}
  \hline
$\lambda$ & \multicolumn{2}{c|}{$L=20$} & \multicolumn{2}{c}{$L=1280$} \\
\hline &  $\chi_1$ & $\chi_2$ & $\chi_1$ & $\chi_2$ \\
  \hline  
0.0  & 10.74(1) & 0 & 5531(13) & 0 \\
0.05  & 11.21(1) & 8.09(1) & 14466(15) & 14464(15)\\
0.10  & 11.72(1) & 10.29(1) & 15839(11) & 15838(11)\\
0.20  & 12.29(0) & 11.59(0) & 16702(15) & 16701(15)\\
\hline
\end{tabular}
\caption{ A comparison of $\chi_1$ and $\chi_2$ as a function of $\lambda$ and $L$.}
 \label{Tab:Ia}
\end{table}
Due to this similarity we only focus on $\chi_1$ in our work.

\section{IV. Plots of $\rho$ and $\chi_1$}
We have computed the fermionic XY model at various values of $\lambda$ on square lattices up to  $L = 4000$ using the
worm algorithm described above. For our simulations, after allowing for appropriate thermalization, we have recorded between $8 \times 10^3$ and $48 \times 10^3$ measurements, each averaged over $2000$ worm updates. A comparable number of measurements were also made for the bosonic model.

In \cref{fig:rho}, we plot $\rho$ for various lattice sizes at different values of $\lambda$ on the left. We note that $\rho$ increases monotonically and approaches the thermodynamic limit by $L=160$ which is shown on the right.

\begin{figure}
 \includegraphics[width=0.45\textwidth]{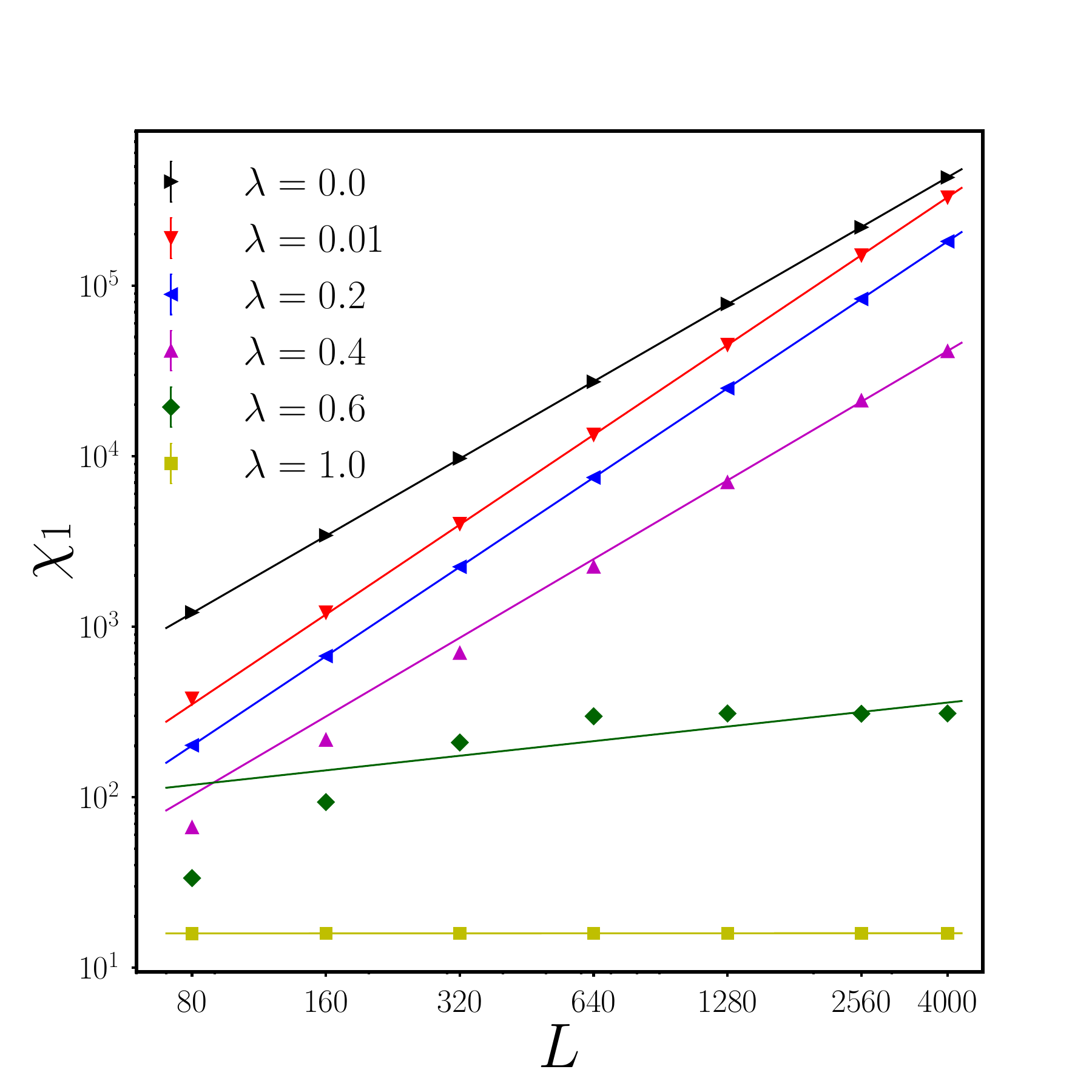}
 \caption{Plot of the finite size scaling for the susceptibility ($\chi_1$) at coupling 
 $\lambda=0.0, 0.01, 0.2, 0.4, 0.6$, and $1.0$. The data (and the corresponding fits) have been vertically shifted to make all the fits in an order. }
 \label{fig:chi}
\end{figure}

In \cref{fig:chi}, we plot $\chi_1$ as a function of system size, $L$ for different values of $\lambda$. When $\lambda$ is small, we find that our data is consistent with the behavior $\chi_1 \sim AL^{2-\eta}$ expected in a critical phase. However, for larger values of $\lambda$, the susceptibility begins to saturate as $\chi_1 \sim A$ which means $\eta \approx 2$. For $\lambda=0$, since the model describes two decoupled layers of close-packed dimer models we expect $\eta=0.5$ \cite{PhysRev.132.1411}. However, when $\lambda$ is small, since we expect our model to describe the physics at the BKT transition, we expect $\eta \sim 0.25$. This is consistent with our findings. 
The values of constant $A$ and $\eta$ for various values of $\lambda$ obtained from a fit are given in \cref{Tab:II}.
 \begin{table}[h]
\renewcommand{\arraystretch}{1.3}
\setlength{\tabcolsep}{12pt}
    \centering
\begin{tabular}{l|cccccc r@{.}l c}
  \hline
$\lambda$ & \multicolumn{1}{c}{$A$} & \multicolumn{1}{c}{$\eta$}  & \multicolumn{1}{c}{${\chi}^2/DOF$}\\
  \hline
0.0   & 0.118(7)   & 0.496(8)   & 1.65 \\
0.01  & 0.041(1)   & 0.250(4)   & 0.62 \\
0.2   & 0.065(1)   & 0.260(3)   & 0.06 \\
0.4   & 0.248(5)   & 0.466(3)   & 399.75 \\
0.6   & 113.13(27) & 1.716(0)   & 70.00 \\
0.7   & 86.75(16)  & 1.876(0)   & 0.28  \\
1.0   & 15.83(2)   & 1.999(0)   & 0.35 \\
  \hline
\end{tabular}
\caption{ Values of constant $A$ and exponent $\eta$ obtained by fitting $\chi_1 = AL^{2-\eta}$ for different 
 coupling values.}
 \label{Tab:II}
\end{table}
 
\section{V. Step Scaling Function}

Let us refer to the traditional $XY$ model defined through the action 
\begin{align}
S = -\beta \sum_{\langle ij \rangle} \cos(\theta_i-\theta_j),
\end{align}
as the bosonic $XY$ model (bXY) and dimer model defined in \cref{eq:2DimerModel} as the fermionic $XY$ model (fXY). In order to argue that these two models are equivalent we compute the universal step scaling function (SSF) in both of them and argue that we get identical results. The SSF is defined as a function between two dimensionless quantities, $\xi(L)/L$ (usually plotted on the $x$ axis) vs. $\xi(2L)/\xi(L)$ (plotted on the $y$ axis). Here we define the finite size correlation length $\xi(L)$ in a finite box of size $L$ using the expression
\begin{align}
\xi(L) = \frac{1}{2\sin{(\pi/L)}} \sqrt{\frac{\chi}{F} - 1},
\end{align}
where 
\begin{align}
\chi &= \sum_i \langle {\cal O}^+_i {\cal O}^-_{0}\rangle,\\
F &= \sum_i \langle {\cal O}^+_i {\cal O}^-_{0}\rangle \cos{(2\pi x /L)},
\end{align}
where $i=(x,t)$ is the space-time lattice site and ${\cal O}^+_i, {\cal O}^-_i$ are lattice fields in the two lattice models. In the bXY model, ${\cal O}^+_i = e^{i\theta_i}$ and ${\cal O}^-_i = e^{-i\theta_i}$, while in the fXY model ${\cal O}^+_i = {\cal O}^-_i = \overline{\psi}_i\psi_i$.

\begin{figure}
 \includegraphics[width=0.5\textwidth]{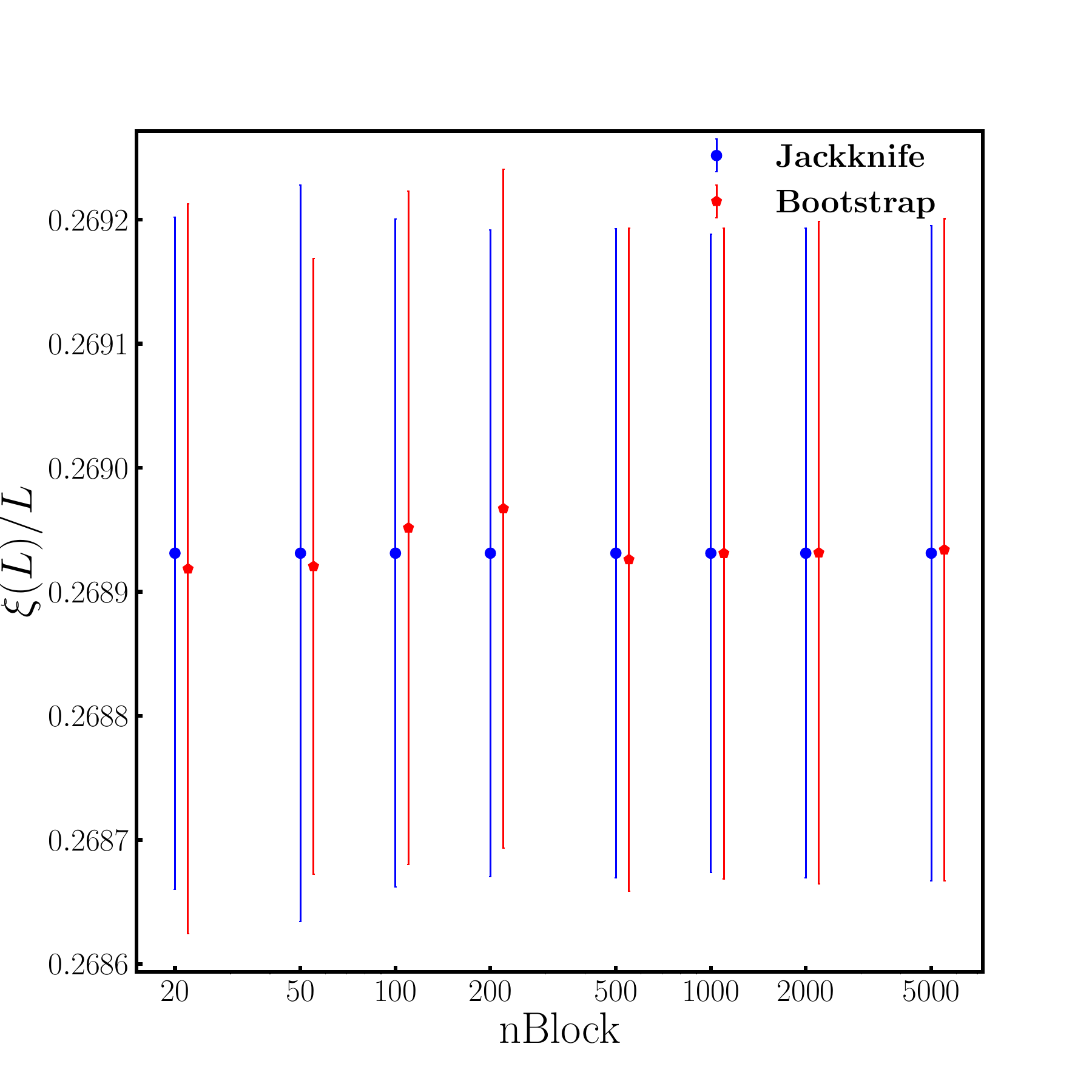}
 \caption{Error analysis of $\xi(L)/L$ at $\lambda=0.65$ and $L=320$ using both the jackknife and the bootstrap method. For the analysis, we have taken $10 \times 10^3$ measurements, each averaged over $2 \times 10^3$ values. In this plot, we used $20,50,100,200,500,1000,2000$ and $5000$ block sizes.}
 \label{fig:u_er}
\end{figure}

We use the jackknife method to compute errors in $\xi(L)/L$ obtained using our Monte Carlo method. Each available data set is divided into 40 jackknife blocks during the error analysis. The effect of varying the jackknife blocks did not change the errors significantly, and the errors were consistent with those obtained using a bootstrap analysis. As an illustration of our analysis, in \cref{fig:u_er} we show an example of the variation of the average and error of $\xi(L)/L$ at $\lambda=0.65$ and $L=320$ for the fermionic model using both the jackknife and the bootstrap analysis as a function of block size. For both methods, we use the same number of block sizes, but in order to show the distinction between them, we have displaced the data on the x-axis by multiplying \emph{nBlock} by a factor of 1.1 for the bootstrap analysis.

\begin{table*}[!htbp]
\renewcommand{\arraystretch}{1.6}
\setlength{\tabcolsep}{14pt}
\begin{tabular}{|c|c||c|c|}
\TopRule
 & \multicolumn{1}{c||}{$bXY$}  &
 & \multicolumn{1}{c|}{$fXY$} \\
\TopRule
$\beta$ & $L$ &
$\lambda$ & $L$ \\
\MidRule
0.92  & 80, 160  & 0.01 & 1280 \\
0.94  & 80, 160  & 0.25 & 1280 \\
0.955 & 80       & 0.35 & 1280 \\
0.975 & 80       & 0.4  & 640, 1280 \\
0.98  & 80       & 0.45 & 1280 \\
0.995 & 80       & 0.5 & 1280 \\
1.015 & 160      & 0.55 & 640, 1280 \\
1.024 & 160, 320 & 0.62 & 80, 160 \\
1.026 & 160      & 0.64 & 80, 160 \\
1.03  & 160      & 0.65 & 160 \\
1.035 & 160      & 0.66 & 160 \\
1.064 & 640      & 0.7 & 160 \\
1.066 & 640      & 0.72 & 160 \\
1.07  & 640      & 0.75 & 160 \\
1.072 & 640      & 0.77 & 160 \\
1.078 & 640      & 0.84 & 160, 320 \\
1.084 & 640      & 0.86 & 160 \\
\BotRule
\end{tabular}
\caption{Table of the pair of ($\beta, L$) and ($\lambda, L$) values used for step-scaling function for both the bosonic and the fermionic $XY$ model. Our goal was to obtain several data points for $L > L_{min}$ as discussed in the text but also separated from each other to as to obtain a smooth curve.}
\label{Tab:XX}
\end{table*}

\begin{table*}[htbp]
\renewcommand{\arraystretch}{1.4}
\setlength{\tabcolsep}{1.9pt}
    \centering
\begin{tabular}{l|ccccc r@{.}l c}
  \hline
Range & \multicolumn{5}{c}{$bXY$} \\
   \hline
$\xi(L)/L$  &  $a_1$ & $a_2$ &  $a_3$ & $a_4$ &  $\chi^2/DOF$\\
  \hline  
 0.066-0.7506912 & 1.74(14) & -9(3) & 236(23) & -648(48) & 0.22 \\
 0.066-0.7506912 & 1.35(4)  & -     & 171(4)  & -517(15) & 0.66 \\
 0.066-0.572  & 1.49(6)  & -     & 135(15) & -321(81) & 0.28 \\
  \hline
  \hline
Range & \multicolumn{5}{c}{$fXY$} \\
   \hline
$\xi(L)/L$   &  $a_1$ & $a_2$ &  $a_3$ & $a_4$ &  $\chi^2/DOF$\\
  \hline  
 0.061-0.7506912 & 1.42(7) & 2(2) & 153(11) & -475(25) & 0.48\\
 0.061-0.7506912 & 1.48(2) &  -   & 165(2)  & -499(8)  & 0.50\\
  \hline
\end{tabular}
\caption{ Values of  $a_1, a_2, a_3, a_4$ and $\chi^2/DOF$ obtained from the fit function $\Sigma(x)$ (defined in \cref{eq:SSF1}) for both the bXY model and the fXY model.}
 \label{Tab:III}
\end{table*}

\begin{figure*}[!htb]
\centering
\hbox{
\includegraphics[width=0.48\textwidth]{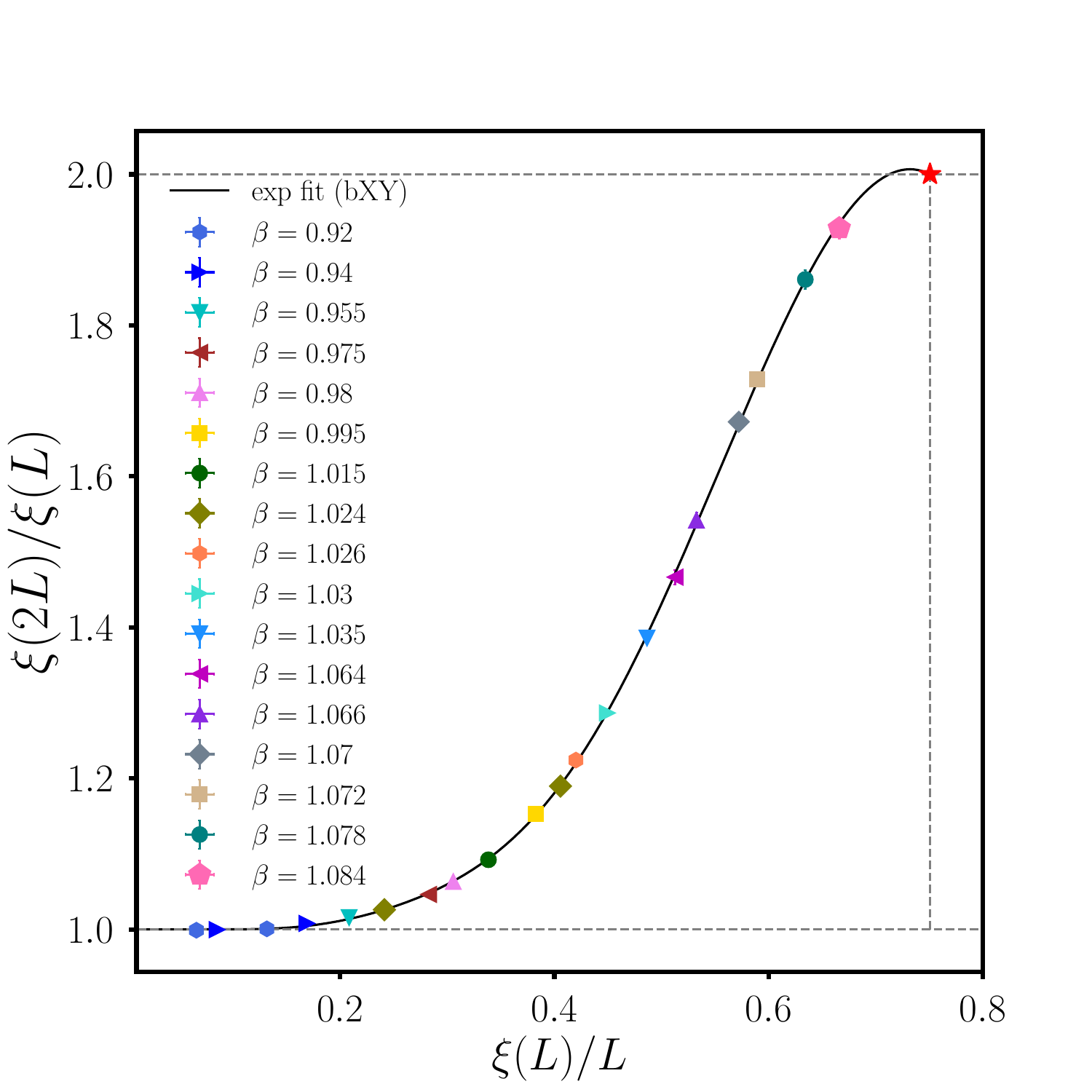}
\includegraphics[width=0.48\textwidth]{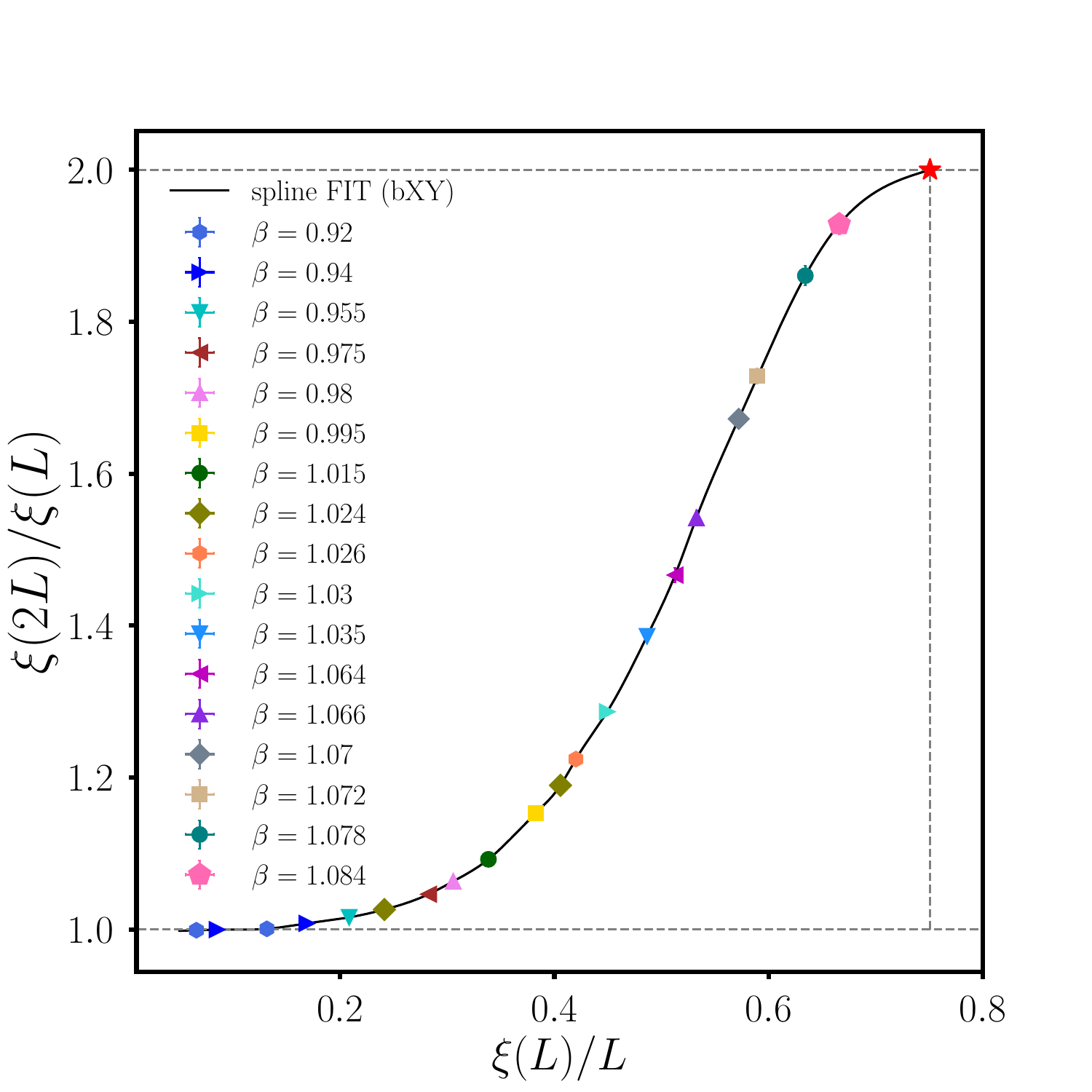}
}
\caption{Plot of the step scaling function for the bosonic XY model superimposed with our data. The data points are from \cref{Tab:XX}. The black line in the left plot is the function \cref{eq:SSF1} with fit parameters from \cref{Tab:III}. In the right plot, the line shows the cubical spline interpolation.}
\label{fig:bXY-SSF}
\end{figure*}

\begin{figure*}[!hbt]
\centering
\hbox{
\includegraphics[width=0.48\textwidth]{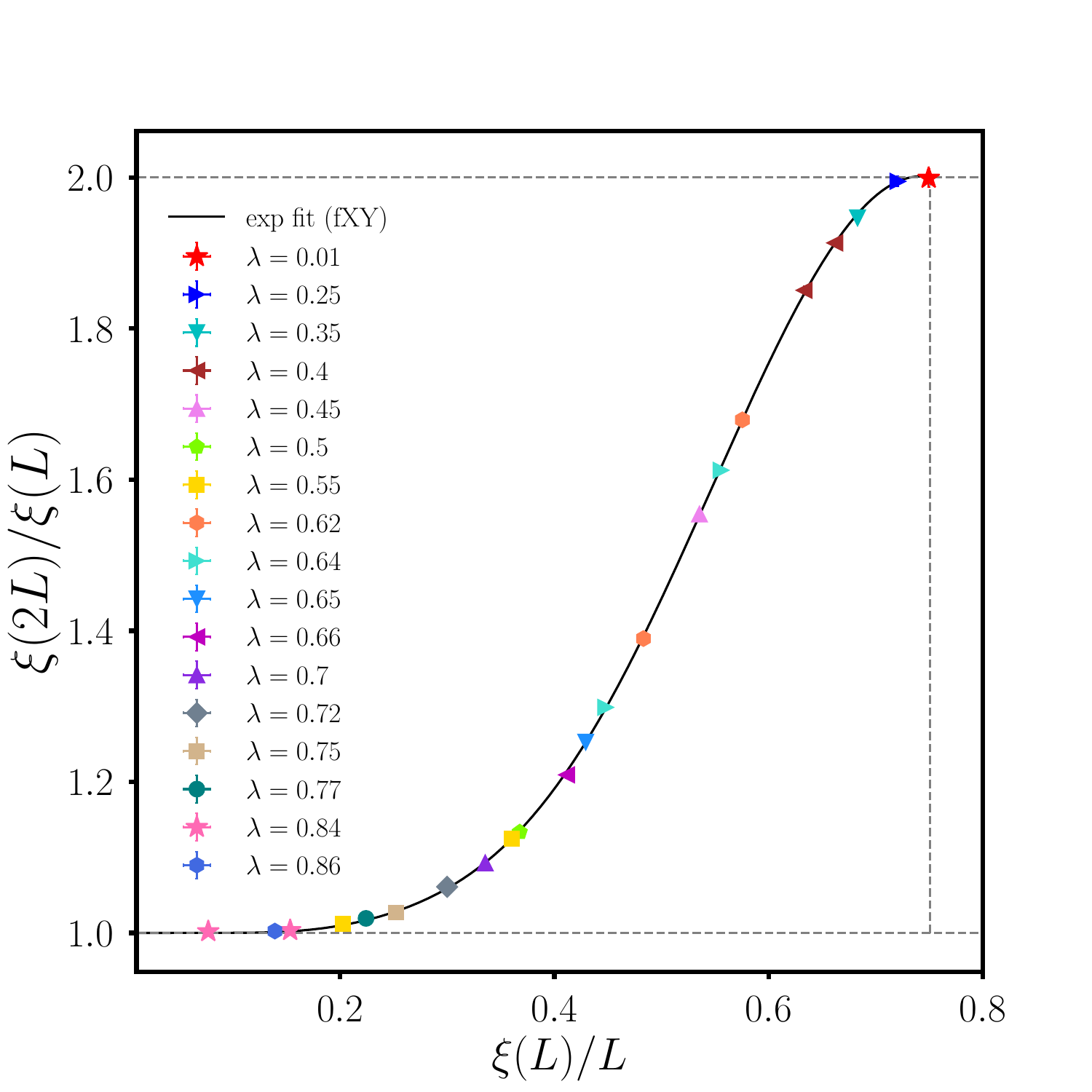}
\includegraphics[width=0.48\textwidth]{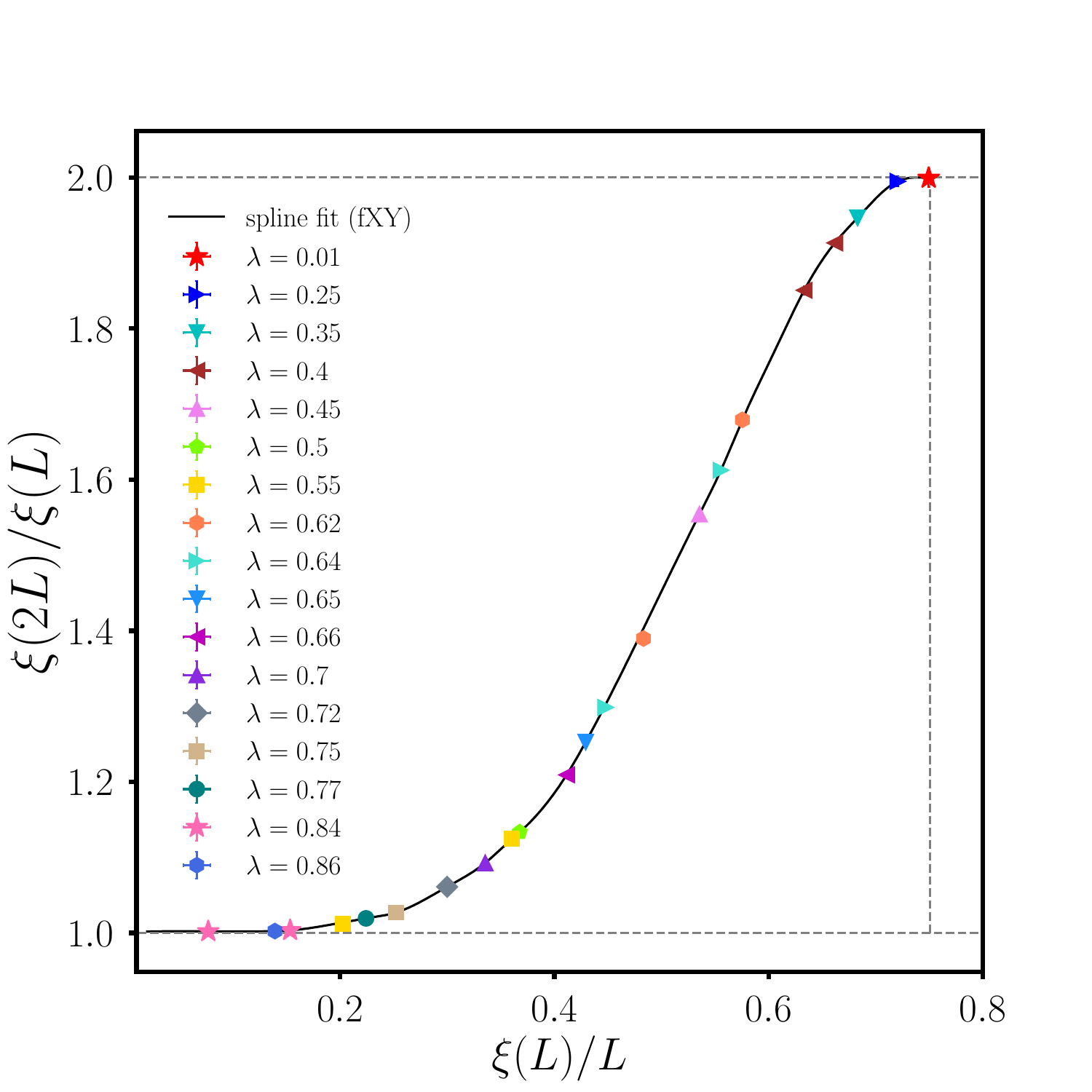}}
\caption{Plot of the step scaling function for the fermionic XY model superimposed with our data. The data points are from \cref{Tab:XX}. The black line in the left plot is the function \cref{eq:SSF1} with fit parameters from \cref{Tab:III}. In the right plot, the line shows the cubical spline interpolation.}
\label{fig:fXY-SSF}
\end{figure*}

\begin{table*}[!htbp]
\renewcommand{\arraystretch}{1.5}
\setlength{\tabcolsep}{15pt}
\begin{tabular}{|c|c|c||c|c|c|}
\TopRule
 & \multicolumn{1}{c|}{$bXY$} &
\multicolumn{1}{c||}{$fXY$} &
 & \multicolumn{1}{c|}{$bXY$} &
\multicolumn{1}{c|}{$fXY$} \\ 
\TopRule
$\xi(L)/L$ & $\xi(2L)/\xi(L)$ & $\xi(2L)/\xi(L)$ &
$\xi(L)/L$ & $\xi(2L)/\xi(L)$ & $\xi(2L)/\xi(L)$ \\
\MidRule
0.05 & 0.998(9)  & 1.000(5) &
0.07 & 0.999(4)  & 1.002(2) \\
0.09 & 0.999(3)  & 1.002(4) &
0.11 & 0.998(5)  & 1.002(4) \\
0.13 & 1.000(4)  & 1.002(4) &
0.15 & 1.004(4)  & 1.003(2) \\
0.17 & 1.008(4)  & 1.006(4) &
0.19 & 1.012(4)  & 1.011(4) \\
0.21 & 1.016(3)  & 1.016(4) &
0.23 & 1.022(4)  & 1.020(3) \\
0.25 & 1.029(5)  & 1.026(1) &
0.27 & 1.039(4)  & 1.037(2) \\
0.29 & 1.051(4)  & 1.053(3) &
0.31 & 1.067(4)  & 1.069(2) \\
0.33 & 1.083(4)  & 1.087(2) &
0.35 & 1.107(5)  & 1.111(4) \\
0.37 & 1.135(6)  & 1.136(4) &
0.39 & 1.163(5)  & 1.167(6) \\
0.41 & 1.199(7)  & 1.206(6) &
0.43  & 1.246(7) & 1.254(3) \\
0.45  & 1.288(7) & 1.308(7) &
0.47  & 1.339(11) & 1.365(9) \\
0.49  & 1.396(9)  & 1.423(11) &
0.51  & 1.457(17) & 1.481(12) \\
0.53  & 1.532(14) & 1.539(8) &
0.55  & 1.602(16) & 1.595(6) \\
0.57  & 1.665(17) & 1.661(8) &
0.59  & 1.729(22) & 1.724(7) \\
0.61  & 1.792(22) & 1.783(7) &
0.63  & 1.849(21) & 1.842(8) \\
0.65  & 1.898(14) & 1.890(5) &
0.67  & 1.935(18) & 1.927(5) \\
0.69  & 1.961(23) & 1.957(9) &
0.71  & 1.977(22) & 1.985(6) \\
0.73  & 1.988(16) & 1.999(6) &
0.7506912  & 2.0 & 1.999(12) \\
\BotRule
\end{tabular}
\caption{The data for the step scaling function obtained via spline fit for both the bosonic and the fermionic $XY$ model. We fix}
 \label{Tab:VI}
\end{table*}

We first compute the SSF for the bXY model since it was not easily available in the literature. For this we vary $\beta$ in the massive phase close to the critical point $\beta_c = 1.1199$ obtained in \cite{Hasenbusch:2005xm}. Our results for $\xi(L)$ are shown in \cref{Tab:XII} and \cref{Tab:XIII}. From this we prepare pairs of data at $(\beta, L)$ and $(\beta,2L)$, and compute both $\xi(2L)/\xi(L)$ and $\xi(L)/L$. Due to finite lattice spacing errors, not all of our data with a given $(\beta, L)$ fall on a universal curve. The lattice spacing errors increase close to $\beta_c$, since we need larger lattices to reach the scaling regime even in the UV. To minimize these errors, as explained in \cite{Caracciolo:1994ud}, we need to choose lattice sizes $L \geq L_{min}$ for each value of $\beta$. We found that for $0.92\leq \beta\leq 0.995$ we needed $L_{min}=80$, for $1.015\leq \beta\leq 1.035$ we needed $L_{min}=160$, and for 
$1.064\leq \beta\leq 1.084$ we needed $L_{min}=640$. 
The final choice of the data we used for extracting the SSF is given in \cref{Tab:XX} and also shown in \cref{fig:bXY-SSF}. Note that we also include by hand the final UV point of the SSF function $\xi(L)/L=0.7506912$ and $\xi(2L)/\xi(L)=2$. We repeated the same procedure for the fXY model. Here we discovered that in order for the data to fall on a universal curve we could use $L_{min} = 80$ for $0.62\leq\lambda\leq0.86$, but had to use $L_{min} = 640$ for $0.25 \leq \lambda \leq 0.6$ and $L_{min} = 1280$ for $\lambda = 0.01$. The choice of our data that we used finally is shown in \cref{Tab:XX} and \cref{fig:fXY-SSF}.

In order to compare the SSF between the bosonic and the fermionic models we tried to parameterize the data shown in \cref{fig:bXY-SSF} and \cref{fig:fXY-SSF} using a function in two different ways. In the first approach, we follow the idea discussed in \cite{Caracciolo:1994ud} where it was proposed that 
\begin{align}
\Sigma(x) = 1 + {a_1} e^{-1/x} + {a_2} e^{-2/x} + {a_3} e^{-3/x} + {a_4} e^{-4/x},
\label{eq:SSF1}
\end{align}
where $x = \xi(L)/L$ and $\Sigma = \xi(2L)/\xi(L)$. The behavior of this function is such that, as $x \rightarrow 0$, the function $\Sigma(x)$ approaches 1. While this function is strictly valid only for small $x$ we find that this form fits our data well. The fit results are given in \cref{Tab:III} and shown as solid lines on the left plots \cref{fig:bXY-SSF} and \cref{fig:fXY-SSF}. We see that while we get good fits by including all four fit parameters, we can also fix $a_2=0$ and still get a good fit. There is an argument given in \cite{Hasenbusch:2005xm} that the largest value of $x$ allowed is $0.7506912...$ shown as a vertical dashed line in the figs. Our data naturally approaches this value, in the bosonic case there is an unphysical bump in solid line around $x\approx 0.75$, which seems like an artifact of the parametrization. To remove this we tried a second parametrization, in which we use 
a cubical spline to interpolate the data. In \cref{Tab:VI}, we provide a tabulation of the spline function that helps parameterize the SSF for both the bosonic and the fermionic models. The errors are again obtained using a jackknife analysis. The spline fit is shown as solid lines in the right plots of \cref{fig:bXY-SSF} and \cref{fig:fXY-SSF}. Note that the bump in the solid line in the bXY case has disappeared in this approach.

\begin{figure*}[!htb]
\centering
\vbox{
\hbox{
\includegraphics[width=0.48\textwidth]{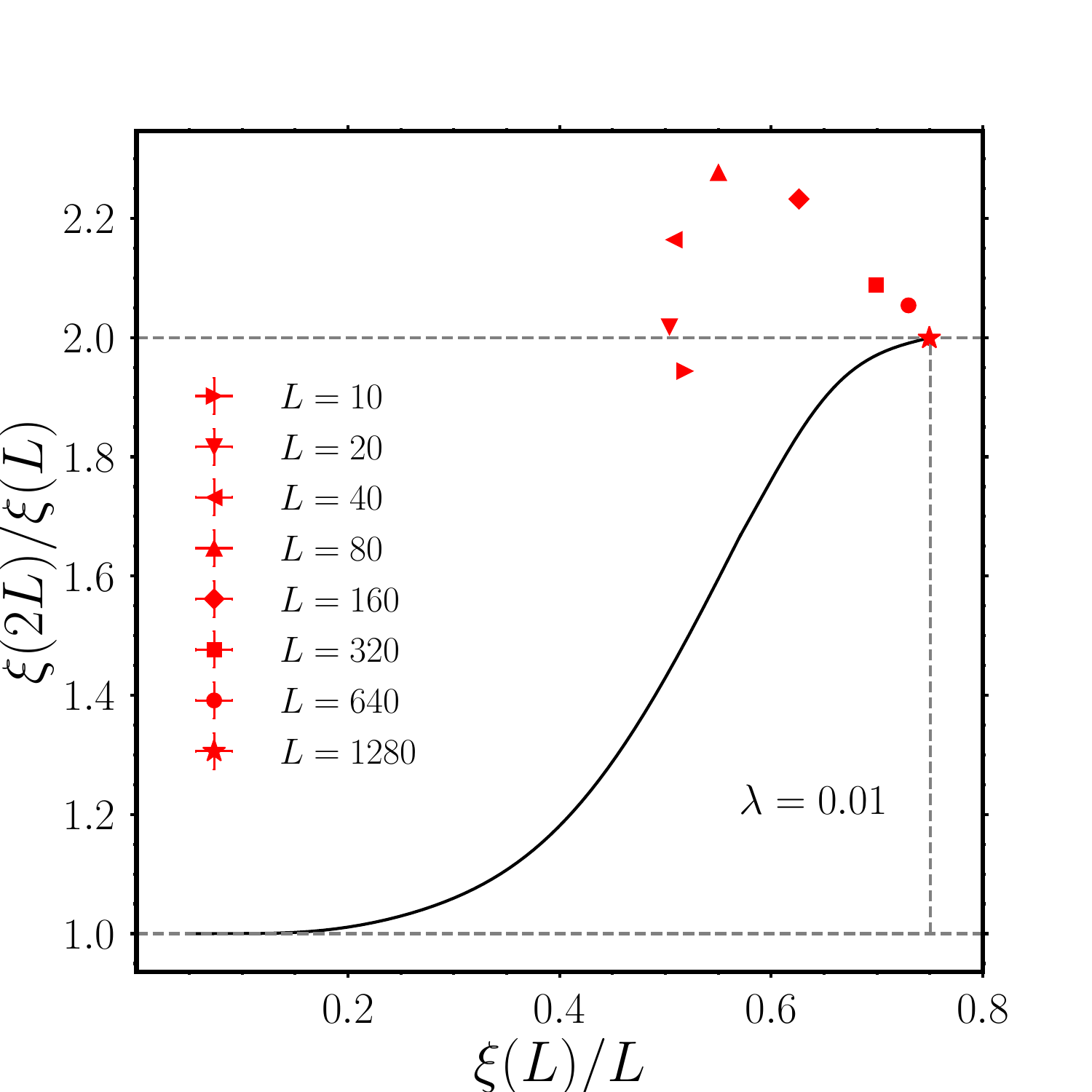}
\includegraphics[width=0.48\textwidth]{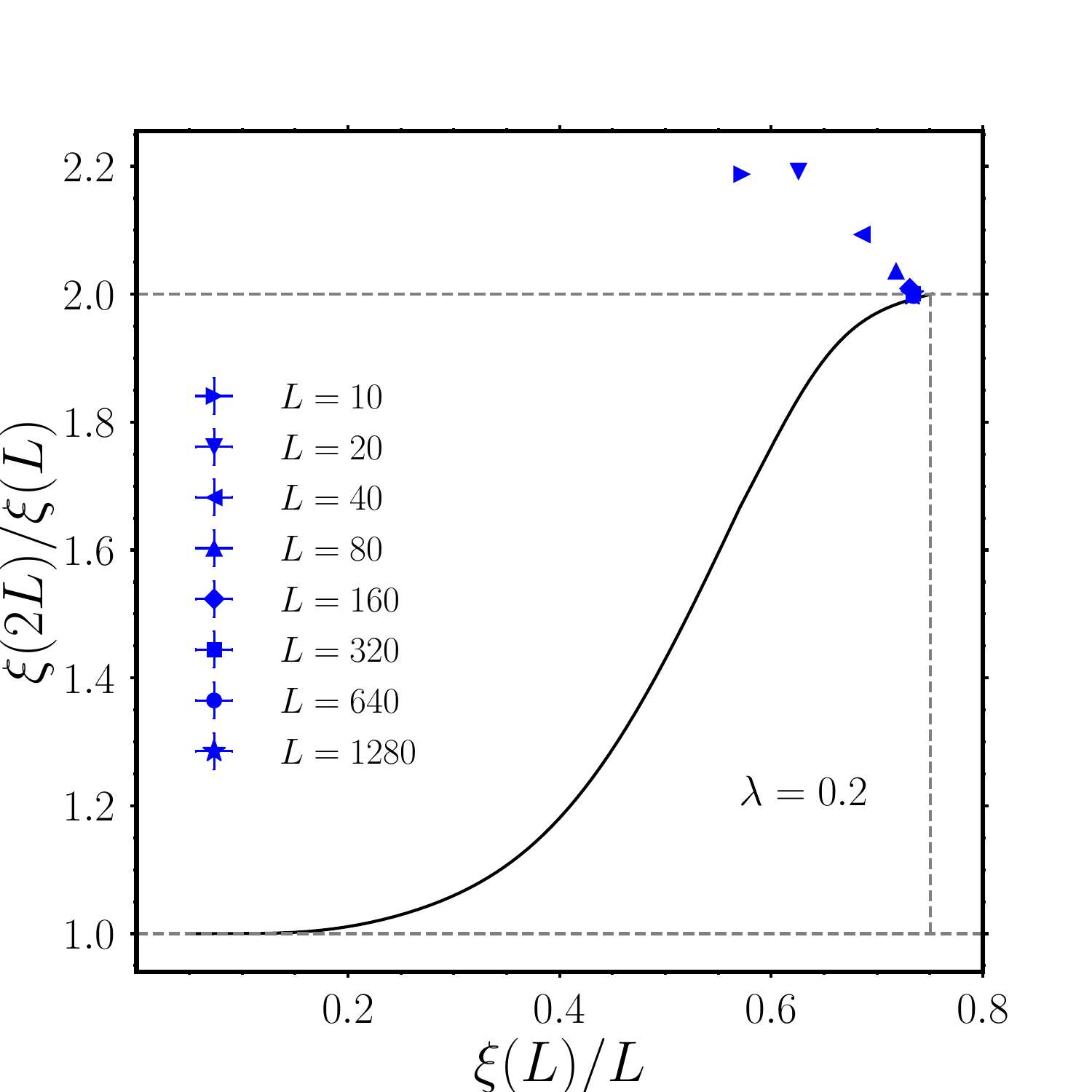}
}
\hbox{
\includegraphics[width=0.48\textwidth]{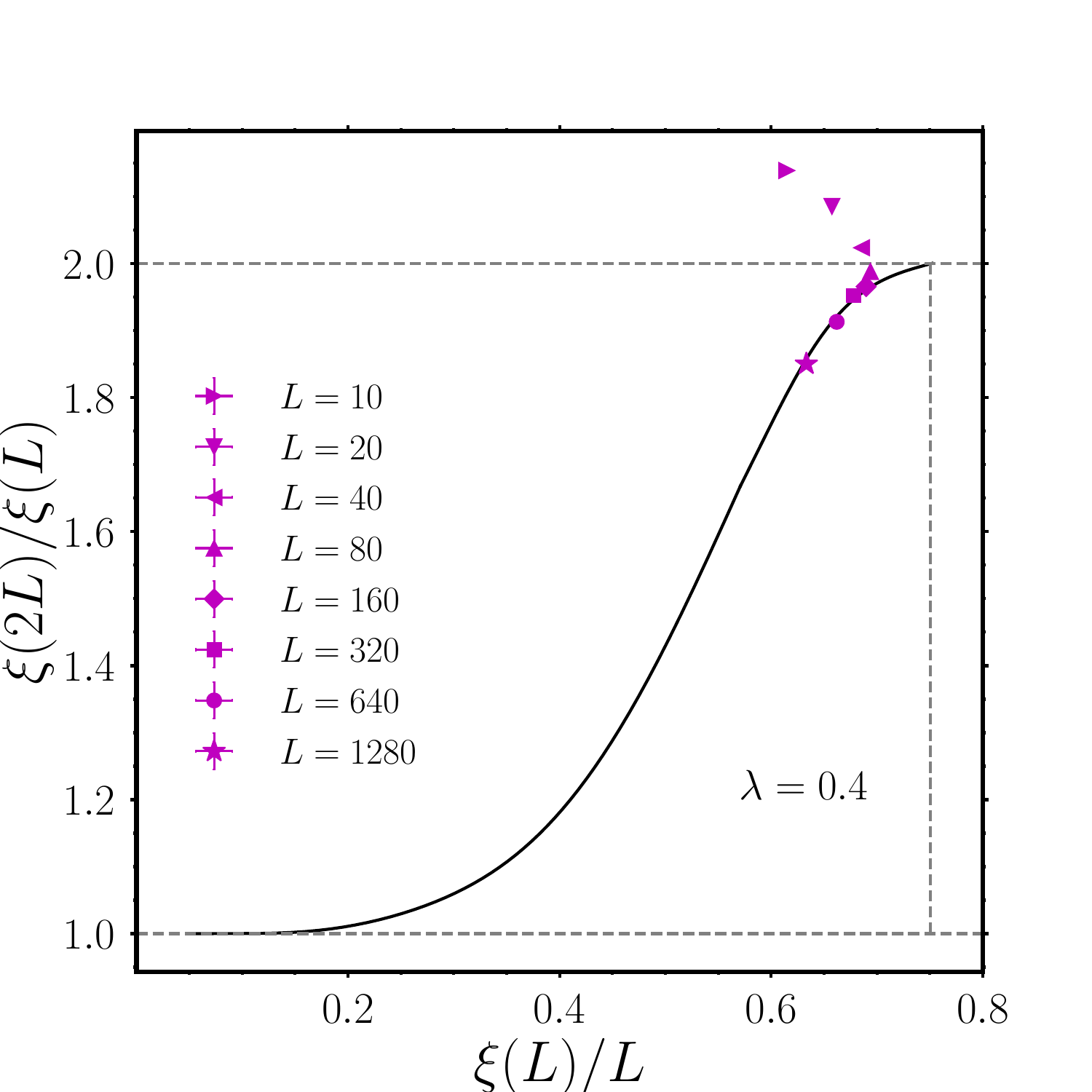}
\includegraphics[width=0.48\textwidth]{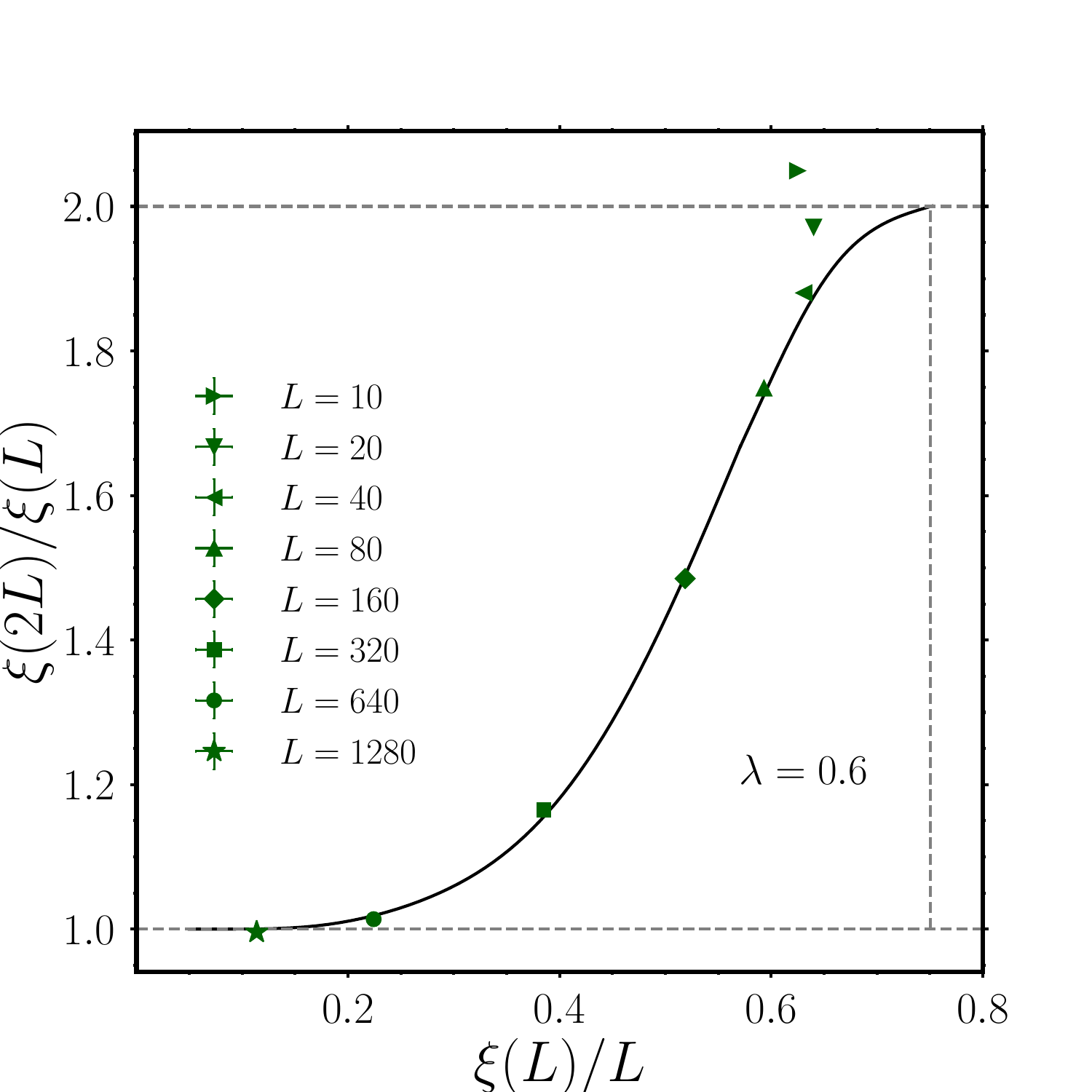}
}}
\caption{Plot of the universal step-scaling function obtained from the bXY model (solid line) and the data from the fXY model at $\lambda = 0.01, 0.2, 0.4$, and $0.6$ for various lattice sizes. This is the data we show in the main paper. We can see that $\Luv \approx 80$ at $\lambda=0.6$, $\Luv \approx 160$ at $\lambda = 0.4$ and $\Luv > 1280$ for $\lambda=0.2$ and $0.01$.}
\label{fig:fXY-SSF-paperdata}
\end{figure*}

\begin{figure}[!hbt]
\centering
\includegraphics[scale=0.35]{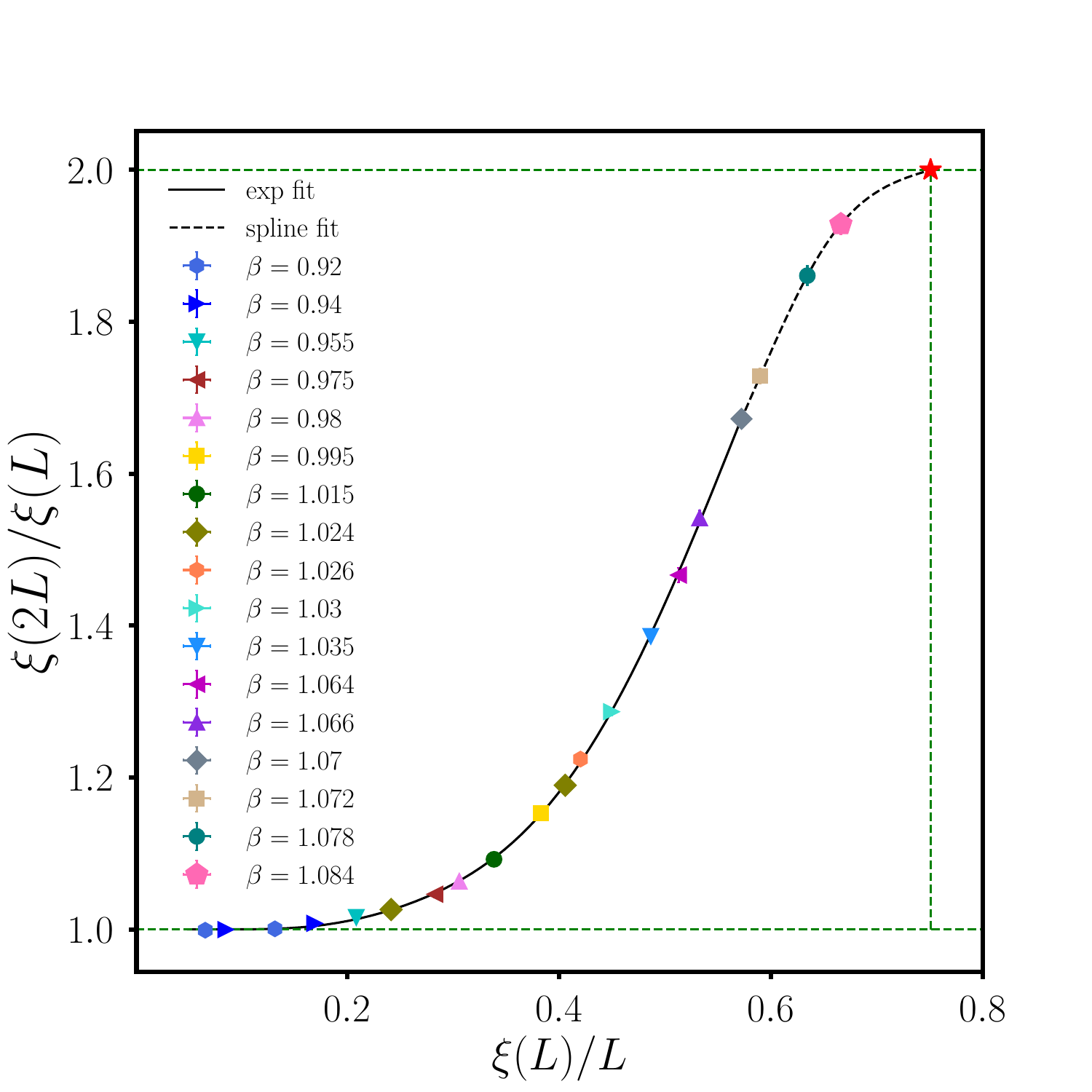}
\caption{Plot of the step scaling function for the bosonic XY model using a combination of the exponential function and the spline function. In particular, we use \cref{eq:SSF1} for $\xi(L)/L \leq 0.572$ (solid line), and the cubical spline interpolation for $\xi(L)/L \geq 0.572$ (dashed line). We use this plot in the main paper.}
\label{fig:bXY-SSF-comb}
\end{figure}

Perhaps the best way to parametrize the true function is to combine the two approaches. Hence, we decided to use \cref{eq:SSF1} for $\xi(L)/L \leq 0.572$ and the cubical spline interpolation for $\xi(L)/L \geq 0.572$. Using this combined form we obtain the solid line shown in \cref{fig:bXY-SSF-comb} for the bXY model, which is also the line we use in the main paper to compare with the our fXY model data. In \cref{fig:fXY-SSF-paperdata} we show the data shown in the main paper, but by separating the various $\lambda$ values for clarification. We notice that the Monte Carlo data for each $\lambda$ do not fall on the solid curve for small values of $L$ but do so for sufficiently large values of $L$. We can define $\Luv$ for each $\lambda$ as the minimum value of $L$ when the data begin to fall on the solid curve. From \cref{fig:fXY-SSF-paperdata} we notice that $\Luv\approx 80$ for $\lambda = 0.6$, and $\Luv\approx 160$ for $\lambda = 0.4$. For $\lambda \leq 0.2$ we notice that $\Luv > 1280$, implying that we will need much larger lattices beyond our current resources to see the data at these couplings to fall on the solid curve. Indeed in  \cref{fig:fXY-SSF-paperdata} we only see the data for these couplings approach the UV prediction of $\xi(L)/L = 0.7506912...$ as explained in  \cite{Hasenbusch:2005xm}.

\section{VI. Infinite Volume Correlation Length}

We can compute the infinite volume correlation length $\xi_\infty$ using the SSF. Here we try to understand how $\xi_\infty$ depends on $\lambda$ in the fermionic $XY$ model. In order to reliably estimate the errors in $\xi_\infty$ we again use the jackknife analysis. We start with 40 jackknife blocks, where each block contains a pair ($\xi(L)/L$, $\xi(2L)/\xi(L)$) for different coupling values ($0.01 \leq \lambda \leq 0.8$). We obtain 40 different cubical splines using each jackknife block. We then start with the initial $\xi(L)/L$ at $L=640$ in each block and evaluate $\xi(2^n L)$ using the spline function for arbitrary values of $n$, until the correlation length $\xi(2^n L)$ becomes insensitive to $L$. Finally, the jackknife mean and error is then computed from the 40 values. These results for $\xi_\infty$ and their errors are quoted in \cref{Tab:IV}.
\begin{table}[h]
\renewcommand{\arraystretch}{1.3}
\setlength{\tabcolsep}{12pt}
    \centering
\begin{tabular}{l|cccccc r@{.}l c}
  \hline
$\lambda$ & \multicolumn{1}{c}{$\xi_{\infty}$}\\
  \hline
0.3   & 145803(94882) \\
0.35  & 18275(1450)   \\
0.4   & 4196(87)      \\
0.45  & 1335(43)      \\
0.5   & 538(6)        \\
0.55  & 262(1)        \\
0.6   & 144.9(5)      \\
0.65  & 89(2)         \\
0.7   & 59.38(23)     \\
0.75  & 41.25(9)      \\
0.8   & 30.54(12)     \\
  \hline
\end{tabular}
\caption{ Values of infinite volume correlation length for different couplings.}
 \label{Tab:IV}
\end{table}
Since the correlation lengths increase exponentially as $\lambda$ becomes small, we were able to extract the infinite volume correlation length only in the range $0.3\leq\lambda\leq0.8$. Below $\lambda < 0.3$, our extrapolation methods fail. 

\begin{figure}
 \includegraphics[width=0.5\textwidth]{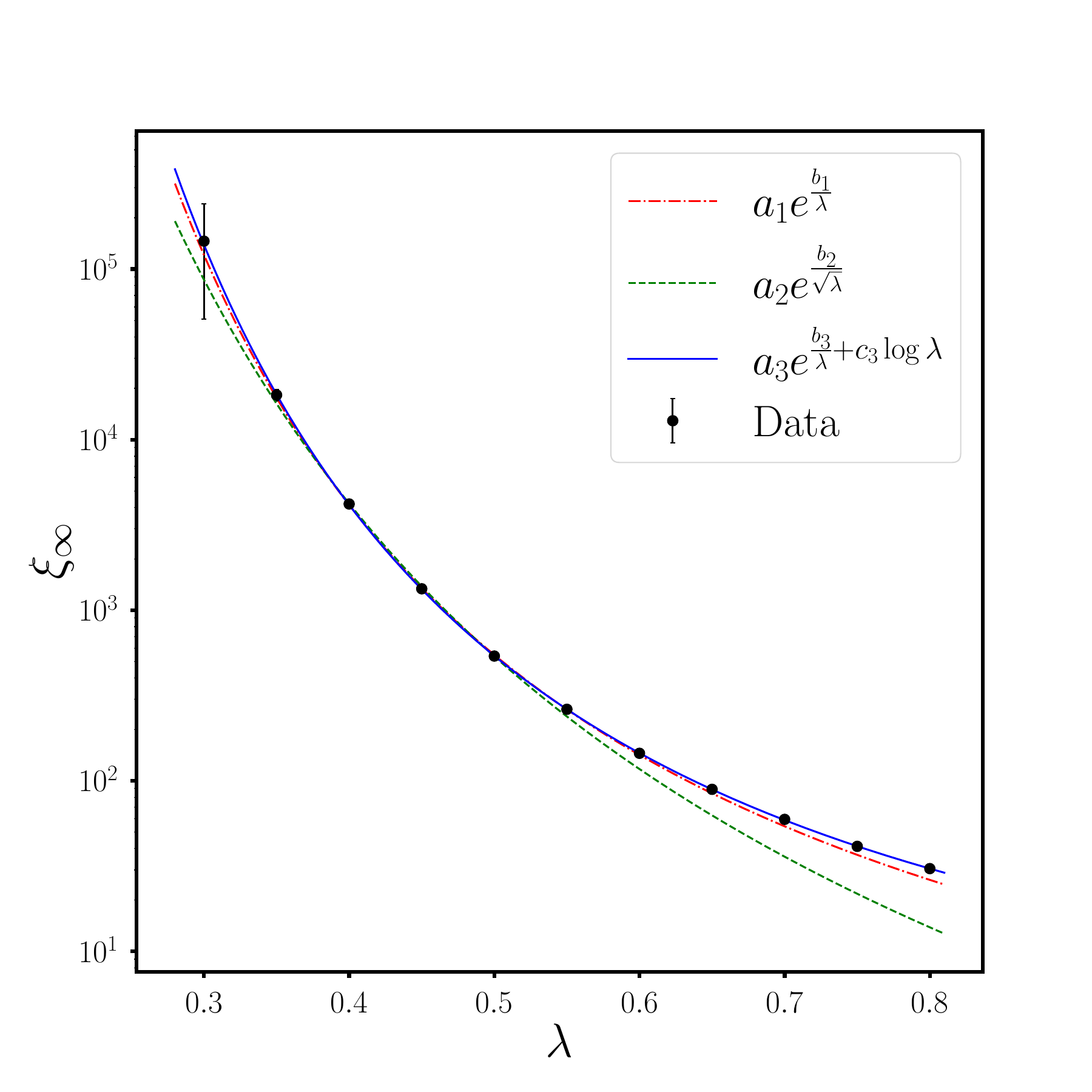}
 \caption{Plot of the infinite volume correlation length with coupling $\lambda$ for the fermionic $XY$ model
 together with several different functional forms. It seems that the correlation length increases
 exponentially with $\frac{1}{\lambda}$.}
 \label{fig:fXY-infCorr}
\end{figure}

Using the data in \cref{Tab:IV} we study the $\lambda$ dependence of $\xi_\infty$. For the bosonic $XY$ model, it is well known that as one approaches the BKT phase transition, the leading divergence of the infinite volume correlation length is captured by
\begin{equation}
\xi = C \exp \left( \frac{b}{\sqrt{\beta_c - \beta}} \right),
\label{eq:xiBXY}
\end{equation}
where $\beta_c$ is the critical coupling, and $b$ and $C$ are non-universal constants. For the fermionic $XY$ model since the partition function is an even function of $\lambda$ we expect $\xi_\infty$ to be a function of $\lambda^2$. Since the BKT critical point appears when $\lambda \rightarrow 0$, we conjecture that
\begin{equation}
\xi^{(1)}_\infty = a_1 \exp \left( \frac{b_1}{\sqrt{\lambda^2}} \right).
\label{eq:xiFXY}
\end{equation}
We test this conjecture numerically by fitting the data in \cref{Tab:IV} to it. We also compare this to other fit forms including $\xi^{(2)}_\infty = a_2 \exp(b_2/(\lambda^2)^{1/4})$ and $\xi^{(3)}_\infty= a_3 \exp(b_3/\sqrt{\lambda^2} + c_3 \log(\lambda^2)/2)$. The results are shown in \cref{Tab:V}. We observe that \cref{eq:xiFXY} is clearly quite good if we expect the constants $a$ and $b$ to be numbers which are not unnatural. We cannot rule out the presence of a power law correction to the expected form.
\begin{table}[h]
\renewcommand{\arraystretch}{1.2}
\centering
\begin{tabular}{l|c|cccc}
  \hline
$i$ & Range & $a_i$ & $b_i$ & $c_i$ & ${\chi}^2/DOF$\\
  \hline
1    & 0.3-0.55 & 0.166(8)   & 4.049(26) & -         &  0.76 \\
  \hline
2    & 0.3-0.5 & 1e-5(3e-6)  & 12.35(13) &  -        &  1.21 \\
  \hline
3    & 0.3-0.8 & 0.119(8)   & 4.68(8)   & 1.36(12)  &   1.15 \\
  \hline
 \end{tabular}
\caption{ Values of $a_1, b_1, c_1, a_2, b_2, a_3, b_3$ and ${\chi}^2/DOF$ obtained from three different fit functions.}
\label{Tab:V}
\end{table}
In \cref{fig:fXY-infCorr}, we show the data in \cref{Tab:IV} and the various fits. The first form is the expected behaviour from \cref{eq:xiFXY}. The second form explores a possible dependence on square-root of $\lambda$ which is clearly unnatural.  Finally the third form allows for a logarithmic correction in the exponential (which is equivalent to including a $1/\lambda$ dependence outside the exponential). We note that in this extended form the data in the larger range  of $0.3 \leq \lambda \leq 0.8$ can be fit.

\section{VII. Monte Carlo Results}
We tabulate all of our Monte Carlo data in \cref{Tab:VII,Tab:VIII,Tab:IX,Tab:X,Tab:XI,Tab:XII,Tab:XIII} for both the bosonic $XY$ and the fermionic $XY$ models, for various values of $L$ and couplings. The errors in these primary quantities have been obtained with 20 jackknife blocks. 

\begin{table*}[!htbp]
\small
\renewcommand{\arraystretch}{1.1}
\setlength{\tabcolsep}{14pt}
    \centering
    \begin{tabular}{|c|c|c|c|c|c|c|}
    \hline
        $\lambda$ & $L$ & $\rho$ & $\chi_1$ & $F$ & $\xi(L)$ & $\Upsilon$ \\
        \hline
        0.00 & 10 & 0 & 3.7294(17) & 0.3318(3) & 5.1774(32) & -      \\
        0.01 & 10 & 0.0001(0) & 3.7308(8)  & 0.3312(3) & 5.1836(24) & -      \\
        0.20 & 10 & 0.0280(0) & 3.8439(12) & 0.2847(3) & 5.7207(38) & -     \\
        0.40 & 10 & 0.0698(0) & 3.8385(12) & 0.2488(3) & 6.1466(40) & -       \\
        0.60 & 10 & 0.1165(0) & 3.6879(14) & 0.2319(2) & 6.2470(39) & -       \\
        \hline
        0.00 & 12 & 0 & 4.9365(18) & 0.4431(6) & 6.1519(53) & -      \\
        0.01 & 12 & 0.0001(0) & 4.9415(15) & 0.4437(3) & 6.1511(21) & -      \\
        0.20 & 12 & 0.0288(0) & 5.2022(16) & 0.3684(5) & 6.9979(55) & -      \\
        0.40 & 12 & 0.0701(0) & 5.2157(14) & 0.3247(4) & 7.4981(52) & -     \\
        0.60 & 12 & 0.1171(0) & 5.0057(21) & 0.3069(3) & 7.5585(50) & -      \\
        \hline
        0.00 & 14 & 0 & 6.2486(26) & 0.5667(6) & 7.1146(50) & -      \\
        0.01 & 14 & 0.0002(0) & 6.2558(23) & 0.5664(6) & 7.1216(43) & -      \\
        0.20 & 14 & 0.0290(0) & 6.7316(31) & 0.4579(6) & 8.3168(61) & -      \\
        0.40 & 14 & 0.0703(0) & 6.7761(23) & 0.4065(6) & 8.8942(74) & -      \\
        0.60 & 14 & 0.1175(0) & 6.4768(21) & 0.3907(5) & 8.8683(58) & -      \\
        \hline
        0.00 & 16 & 0 & 7.6612(35) & 0.7005(10) & 8.0789(72)  & -      \\
        0.01 & 16 & 0.0002(0) & 7.6729(30) & 0.6975(8)  & 8.1047(62)  & -      \\
        0.20 & 16 & 0.0292(0) & 8.4206(32) & 0.5513(6)  & 9.6832(60)  & -      \\
        0.40 & 16 & 0.0704(0) & 8.5034(29) & 0.4948(8)  & 10.3112(88) & -     \\
        0.60 & 16 & 0.1177(0) & 8.0952(34) & 0.4834(7)  & 10.1705(84) & -      \\        
        \hline 
        0.00 & 20 & 0 & 10.7423(47) & 0.9926(11) & 10.0173(69)  & 0.6058(2)      \\
        0.01 & 20 & 0.0002(0) & 10.7770(53) & 0.9853(9)  & 10.0758(54)  & 0.6071(4)      \\
        0.05 & 20 & 0.0044(0) & 11.2088(48) & 0.9162(8)  & 10.7130(62)  & 0.6203(2)      \\
        0.10 & 20 & 0.0119(0) & 11.7235(59) & 0.8370(12) & 11.5272(109) & 0.6255(7)      \\
        0.20 & 20 & 0.0293(0) & 12.2905(37) & 0.7525(11) & 12.516(11)   & 0.6172(4)      \\
        0.30 & 20 & 0.0293(0) & 12.2905(37) & 0.7117(12) & 12.9957(132) & 0.6002(4)      \\
        0.40 & 20 & 0.0705(0) & 12.4270(48) & 0.6934(8)  & 13.1483(85)  & 0.5792(4)      \\
        0.50 & 20 & 0.0935(0) & 12.1726(47) & 0.6863(10) & 13.0776(101) & 0.5518(4)      \\
        0.60 & 20 & 0.1180(0) & 11.7508(49) & 0.6895(9)  & 12.8022(94)  & 0.5131(5)      \\   
        0.70 & 20 & 0.1441(0) & 11.1045(46) & 0.7078(8)  & 12.2495(85)  & 0.4660(3)      \\
        0.80 & 20 & 0.1718(0) & 10.2291(46) & 0.7427(8)  & 11.4230(71)  & 0.4039(3)      \\
        1.00 & 20 & 0.2330(0) & 7.7504(38)  & -          & -            & 0.2456(3)      \\
        \hline
        0.00 & 22 & 0 & 12.4079(69) & 1.1521(15) & 10.9814(93)  & -      \\
        0.01 & 22 & 0.0003(0) & 12.4556(56) & 1.1440(14) & 11.0478(80)  & -      \\
        0.20 & 22 & 0.0293(0) & 14.4529(61) & 0.8588(13) & 13.978(13)   & -      \\
        0.40 & 22 & 0.0705(0) & 14.6262(56) & 0.8008(14) & 14.5986(142) & -      \\
        0.60 & 22 & 0.1182(0) & 13.7526(67) & 0.8092(8)  & 14.0516(94)  & -      \\        
        \hline
        0.00 & 24 & 0 & 14.1467(70) & 1.3157(13) & 11.9626(81)  & -      \\
        0.01 & 24 & 0.0003(0) & 14.2197(72) & 1.3055(14) & 12.0483(81)  & -      \\
        0.20 & 24 & 0.0293(0) & 16.7707(98) & 0.9749(14) & 15.419(12)   & -      \\
        0.40 & 24 & 0.0705(0) & 16.9713(77) & 0.9167(10) & 16.031(12)   & -      \\
        0.60 & 24 & 0.1182(0) & 15.8922(78) & 0.9313(14) & 15.354(12)   & -      \\        
        \hline
        0.00 & 26 & 0 & 15.9548(73)  & 1.4888(18) & 12.9302(96)  & -      \\
        0.01 & 26 & 0.0003(0) & 16.0628(85)  & 1.4750(17) & 13.0452(92)  & -      \\
        0.20 & 26 & 0.0293(0) & 19.2352(75)  & 1.0919(12) & 16.909(10)   & -      \\
        0.40 & 26 & 0.0706(0) & 19.4497(68)  & 1.0370(16) & 17.479(15)   & -      \\
        0.60 & 26 & 0.1183(0) & 18.149(11) & 1.0657(13) & 16.608(13)     & -      \\
        \hline        
        0.00 & 28 & 0 & 17.844(11) & 1.6689(22) & 13.903(13) & -      \\
        0.01 & 28 & 0.0003(0) & 17.996(7)  & 1.6558(17) & 14.029(10) & -      \\
        0.20 & 28 & 0.0293(0) & 21.841(11) & 1.2127(24) & 18.418(18) & -      \\
        0.40 & 28 & 0.0706(0) & 22.090(12) & 1.1668(18) & 18.910(17) & -      \\
        0.60 & 28 & 0.1184(0) & 20.527(9)  & 1.2049(19) & 17.883(17) & -      \\
        \hline           
    \hline
    \end{tabular}
    \caption{Results for $\rho$, $\chi_1$, $F$, $\chi(L)$ and $\Upsilon$ for the fermionic $XY$ model, 
    $L=10$, $L=12$, $L=14$, $L=16$, $L=20$, $L=22$, $L=24$, $L=26$,  and $L=28$.} 
    \label{Tab:VII}
\end{table*}

\begin{table*}[!htbp]
\small
\renewcommand{\arraystretch}{1.0}
\setlength{\tabcolsep}{14pt}
    \centering
    \begin{tabular}{|c|c|c|c|c|c|c|}
    \hline
        $\lambda$ & $L$ & $\rho$ & $\chi_1$ & $F$ & $\xi(L)$ & $\Upsilon$ \\
        \hline      
        0.00 & 30 & 0 & 19.814(10) & 1.8528(26) & 14.893(13) & -      \\
        0.01 & 30 & 0.0004(0) & 20.007(7)  & 1.8314(25) & 15.069(13) & -      \\
        0.20 & 30 & 0.0293(0) & 24.612(13) & 1.3429(21) & 19.911(17) & -      \\
        0.40 & 30 & 0.0706(0) & 24.857(10) & 1.3046(19) & 20.324(18) & -      \\
        0.60 & 30 & 0.1184(0) & 22.991(10) & 1.3553(16) & 19.112(13) & -      \\
        \hline
        0.00 & 40 & 0 & 30.528(14) & 2.883(4) & 19.734(17) & 0.6070(2)      \\
        0.01 & 40 & 0.0004(0) & 31.221(15) & 2.792(5) & 20.335(22) & 0.6141(5)      \\
        0.05 & 40 & 0.0049(0) & 35.155(23) & 2.364(3) & 23.732(20) & 0.6326(3)      \\
        0.10 & 40 & 0.0120(0) & 37.847(16) & 2.167(4) & 25.861(27) & 0.6290(7)      \\
        0.20 & 40 & 0.0293(0) & 40.408(23) & 2.068(4) & 27.441(33) & 0.6169(8)      \\
        0.30 & 40 & 0.0491(0) & 41.066(17) & 2.060(5) & 27.728(39) & 0.5990(4)      \\
        0.40 & 40 & 0.0707(0) & 40.678(17) & 2.084(4) & 27.425(32) & 0.5731(5)      \\
        0.50 & 40 & 0.0938(0) & 39.311(18) & 2.124(4) & 26.668(24) & 0.5354(6)      \\
        0.60 & 40 & 0.1186(0) & 36.820(20) & 2.207(3) & 25.239(22) & 0.4787(4)       \\
        0.70 & 40 & 0.1451(0) & 33.059(15) & 2.379(4) & 22.885(21) & 0.3998(4)       \\
        0.80 & 40 & 0.1736(0) & 27.522(12) & 2.644(3) & 19.548(14) & 0.2902(4)       \\
        1.00 & 40 & 0.2368(0) & 13.879(11) & -        & -          & 0.0741(2)       \\
        \hline          
        0.00 & 50 & 0 & 42.748(34) & 4.054(7) & 24.600(24) & -      \\
        0.01 & 50 & 0.0005(0) & 42.381(38) & 3.849(8) & 25.841(34) & -      \\
        0.20 & 50 & 0.0293(0) & 59.431(31) & 2.929(5) & 34.972(36) & -      \\
        0.40 & 50 & 0.0707(0) & 59.612(25) & 3.007(5) & 34.547(31) & -      \\
        0.60 & 50 & 0.1187(0) & 52.929(30) & 3.262(4) & 31.071(23) & -      \\
        \hline           
        0.00 & 60 & 0 & 56.226(41) & 5.356(9) & 29.442(30) & -      \\
        0.01 & 60 & 0.0005(0) & 59.309(35) & 4.957(9) & 31.637(33) & -      \\
        0.20 & 60 & 0.0293(0) & 81.534(35) & 3.922(8) & 42.502(50) & -      \\
        0.40 & 60 & 0.0707(0) & 81.416(54) & 4.078(8) & 41.603(49) & -      \\
        0.60 & 60 & 0.1188(0) & 70.941(38) & 4.479(7) & 36.801(30) & -      \\
        \hline          
        0.00 & 70 & 0 & 70.843(58)  & 6.767(11)  & 34.294(35)  & -      \\
        0.01 & 70 & 0.0006(0) & 76.153(55)  & 6.112(11)  & 37.728(47)  & -      \\
        0.20 & 70 & 0.0293(0) & 106.713(53) & 5.023(11)  & 50.142(59)  & -      \\
        0.40 & 70 & 0.0707(0) & 105.919(52) & 5.315(12)  & 48.486(64)  & -      \\
        0.60 & 70 & 0.1188(0) & 90.612(46)  & 5.915(10)  & 42.170(43)  & -      \\        
        \hline            
        0.00 & 80 & 0 & 86.610(70)  & 8.247(17)  & 39.257(44)  & 0.6067(3)      \\
        0.01 & 80 & 0.0006(0) & 94.754(58)  & 7.322(16)  & 44.010(58)  & 0.6299(4)      \\
        0.05 & 80 & 0.0049(0) & 114.924(74) & 6.133(10)  & 53.637(51)  & 0.6336(4)      \\
        0.10 & 80 & 0.0120(0) & 125.571(77) & 6.104(15)  & 56.343(76)  & 0.6308(8)      \\
        0.20 & 80 & 0.0293(0) & 134.390(89) & 6.296(12)  & 57.445(63)  & 0.6149(7)      \\
        0.30 & 80 & 0.0491(0) & 136.021(90) & 6.476(15)  & 56.960(73)  & 0.5963(5)      \\
        0.40 & 80 & 0.0707(0) & 133.220(76) & 6.663(17)  & 55.505(75)  & 0.5639(7)      \\
        0.50 & 80 & 0.0940(0) & 125.636(80) & 6.959(13)  & 52.593(54)  & 0.5111(7)      \\
        0.60 & 80 & 0.1188(0) & 111.685(68) & 7.503(12)  & 47.457(43)  & 0.4239(6)      \\   
        0.62 & 80 & -         & 107.98(14)  & 7.67(3)    & 46.04(10)   & -              \\
        0.64 & 80 & -         & 103.64(13)  & 7.86(2)    & 44.45(8)    & -              \\
        0.70 & 80 & 0.1457(0) & 88.174(49)  & 8.618(10)  & 38.696(31)  & 0.2825(5)      \\
        0.80 & 80 & 0.1747(0) & 56.099(43)  & 9.908(11)  & 27.499(22)  & 0.1190(3)      \\
        0.90 & 80 & -         & 29.72(4)    & 9.74(1)    & 18.24(2)    & -              \\ 
        1.00 & 80 & 0.2376(0) & 15.851(15)  & -          & -           & 0.0039(1)      \\
        \hline        
        0.00 & 90 & 0 & 103.393(87)  & 9.829(19) & 44.203(55)  & -      \\
        0.01 & 90 & 0.0006(0) & 115.008(71)  & 8.579(24) & 50.461(82)  & -      \\
        0.20 & 90 & 0.0293(0) & 164.845(99)  & 7.676(19) & 64.827(87)  & -      \\
        0.40 & 90 & 0.0707(0) & 162.88(10)   & 8.176(12) & 62.322(60)  & -      \\
        0.60 & 90 & 0.1189(0) & 133.923(72)  & 9.304(18) & 52.433(52)  & -      \\        
        \hline          
        0.00 & 100 & 0 & 121.10(13) & 11.560(26) & 49.001(64)  & -      \\
        0.01 & 100 & 0.0006(0) & 137.33(9)  & 9.822(17)  & 57.354(55)  & -      \\
        0.20 & 100 & 0.0293(0) & 197.89(14) & 9.121(22)  & 72.42(10)   & -      \\
        0.40 & 100 & 0.0707(0) & 194.81(10) & 9.774(19)  & 69.260(75)  & -      \\
        0.60 & 100 & 0.1189(0) & 157.38(13) & 11.262(15) & 57.337(57)  & -      \\        
        \hline           
    \hline
    \end{tabular}
    \caption{Results for $\rho$, $\chi_1$, $F$, $\chi(L)$ and $\Upsilon$ for the fermionic $XY$ model, $L=30$,
    $L=40$, $L=50$, $L=60$, $L=70$, $L=80$, $L=90$,  and $L=100$.}
    \label{Tab:VIII}
\end{table*}

\begin{table*}[!htbp]
\small
\renewcommand{\arraystretch}{1.1}
\setlength{\tabcolsep}{14pt}
    \centering
    \begin{tabular}{|c|c|c|c|c|c|c|}
    \hline
        $\lambda$ & $L$ & $\rho$ & $\chi_1$ & $F$ & $\xi(L)$ & $\Upsilon$ \\
        \hline           
        0.00 & 110 & 0 & 139.51(12)  & 13.313(40) & 53.91(10)  & -      \\
        0.01 & 110 & 0.0006(0) & 161.08(12)  & 11.108(28) & 64.338(96) & -      \\
        0.20 & 110 & 0.0293(0) & 233.79(15)  & 10.739(33) & 79.80(13)  & -      \\
        0.40 & 110 & 0.0707(0) & 229.27(15)  & 11.500(19) & 76.194(75) & -      \\
        0.60 & 110 & 0.1189(0) & 181.84(13)  & 13.404(23) & 62.070(70) & -      \\
        \hline          
        0.00 & 120 & 0 & 159.43(13) & 15.189(32) & 58.862(82)  & -      \\
        0.01 & 120 & 0.0006(0) & 186.01(17) & 12.496(33) & 71.18(12)   & -      \\
        0.20 & 120 & 0.0293(0) & 271.73(14) & 12.447(37) & 87.18(14)   & -      \\
        0.40 & 120 & 0.0707(0) & 265.92(20) & 13.392(21) & 82.945(84)  & -      \\
        0.60 & 120 & 0.1189(0) & 206.95(12) & 15.726(23) & 66.604(59)  & -      \\
        \hline           
        0.00 & 130 & 0 & 179.47(19) & 17.163(50) & 63.63(12)  & -      \\
        0.01 & 130 & 0.0006(0) & 213.22(15) & 13.849(46) & 78.51(14)  & -      \\
        0.20 & 130 & 0.0293(0) & 312.66(17) & 14.241(34) & 94.72(18)  & -      \\
        0.40 & 130 & 0.0707(0) & 304.73(17) & 15.346(32) & 89.85(11)  & -      \\
        0.60 & 130 & 0.1189(0) & 231.93(20) & 18.324(26) & 70.647(75) & -      \\
        \hline
        0.00 & 140 & 0 & 200.42(17) & 19.13(5) & 68.59(12)   & -      \\
        0.01 & 140 & 0.0006(0) & 241.54(22) & 15.35(4) & 85.53(12)   & -      \\
        0.20 & 140 & 0.0293(0) & 355.34(24) & 16.11(4) & 102.24(12)  & -      \\
        0.40 & 140 & 0.0707(0) & 346.00(18) & 17.43(4) & 96.74(12)   & -       \\
        0.60 & 140 & 0.1190(0) & 258.59(22) & 20.90(3) & 75.16(7)    & -      \\
        \hline         
        0.00 & 150 & 0 & 222.47(26) & 21.29(5) & 73.40(10)   & -      \\
        0.01 & 150 & 0.0006(0) & 271.68(28) & 16.82(5) & 92.93(17)   & -      \\
        0.20 & 150 & 0.0293(0) & 400.88(24) & 18.14(5) & 109.68(19)  & -      \\
        0.40 & 150 & 0.0707(0) & 389.17(20) & 19.64(3) & 103.57(09)  & -      \\
        0.60 & 150 & 0.1190(0) & 285.22(24) & 23.86(3) & 79.02(08)   & -     \\
        \hline          
        0.00 & 160 & 0 & 245.14(26) & 23.54(6) & 78.14(12)   & 0.6065(5)  \\
        0.01 & 160 & 0.0006(0) & 302.95(28) & 18.38(7) & 100.20(22)  & 0.6353(6)  \\
        0.05 & 160 & 0.0049(0) & 382.42(29) & 17.88(5) & 114.99(20)  & 0.6334(4)  \\
        0.10 & 160 & 0.0120(0) & 419.42(25) & 18.91(5) & 117.20(18)  & 0.6301(8)  \\
        0.20 & 160 & 0.0293(0) & 448.38(28) & 20.30(4) & 116.96(14)  & 0.6148(10) \\
        0.30 & 160 & 0.0491(0) & 450.15(30) & 21.20(6) & 114.56(19)  & 0.5913(6)  \\
        0.40 & 160 & 0.0707(0) & 434.43(30) & 21.96(5) & 110.36(14)  & 0.5512(7)  \\
        0.50 & 160 & 0.0940(0) & 393.80(17) & 23.35(4) & 101.44(10)  & 0.4753(7)  \\
        0.60 & 160 & 0.1190(0) & 312.00(21) & 26.86(5) & 82.98(9)    & 0.3255(5)  \\
        0.62 & 160 & -         & 287.38(24) & 28.13(8) & 77.31(12)   & -          \\
        0.64 & 160 & -         & 261.76(26) & 29.35(9) & 71.67(12)   & -          \\
        0.65 & 160 & -         & 247.46(18) & 29.90(4) & 68.70(7)    & -          \\                
        0.66 & 160 & -         & 233.07(49) & 30.33(10)& 65.85(16)   & -          \\
        0.70 & 160 & 0.1460(0) & 175.85(13) & 32.30(3) & 53.69(5)    & 0.1111(5)  \\
        0.72 & 160 & -         & 148.60(16) & 32.65(8) & 47.99(7)    & -          \\
        0.75 & 160 & -         & 113.48(11) & 32.33(3) & 40.35(3)    & -          \\
        0.77 & 160 & -         & 93.93(18)  & 31.47(7) & 35.88(5)    & -          \\
        0.80 & 160 & 0.1750(0) & 71.25(7)   & 29.42(3) & 30.37(3)    & 0.0116(2)  \\
        0.84 & 160 & -         & 49.83(10)  & 25.83(4) & 24.55(3)    & -          \\
        0.86 & 160 & -         & 42.23(7)   & 23.93(3) & 22.27(4)    & -          \\
        0.90 & 160 & -         & 30.91(7)   & 20.26(3) & 18.46(3)    & -          \\
        1.00 & 160 & 0.2377(0) & 15.92(1)   & -        & -           & 0  \\
        \hline         
        0.00 & 170 & 0 & 268.05(25)& 25.71(6) & 83.07(11)   & -      \\
        0.01 & 170 & 0.0006(0) & 336.4(2)  & 19.9(1)  & 107.92(24)  & -      \\
        0.20 & 170 & 0.0293(0) & 498.2(3)  & 22.6(1)  & 124.17(20)  & -      \\
        0.40 & 170 & 0.0707(0) & 480.9(3)  & 24.4(1)  & 117.08(15)  & -      \\
        0.60 & 170 & 0.1190(0) & 338.7(2)  & 30.3(1)  & 86.38(11)   & -      \\        
        \hline           
        0.00 & 240 & 0 & 450.53(58)  & 43.17(12)  & 117.34(19)  & -      \\
        0.01 & 240 & 0.0006(0) & 608.83(56)  & 32.19(10)  & 161.66(32)  & -      \\
        0.20 & 240 & 0.0293(0) & 908.26(71)  & 40.96(13)  & 175.78(30)  & -      \\
        0.40 & 240 & 0.0707(0) & 863.59(58)  & 44.60(96)  & 163.69(21)  & -      \\
        0.60 & 240 & 0.1190(0) & 520.76(45)  & 58.57(80)  & 107.31(13)  & -      \\            
     \hline
    \end{tabular}
    \caption{Results for $\rho$, $\chi_1$, $F$, $\chi(L)$ and $\Upsilon$ for the fermionic $XY$ model, $L=110$, $L=120$, $L=130$,
    $L=140$, $L=150$, $L=160$, $L=170$, and $L=240$.} 
    \label{Tab:IX}
\end{table*}

\begin{table*}[!htbp]
\small
\renewcommand{\arraystretch}{1.3}
\setlength{\tabcolsep}{14pt}
    \centering
    \begin{tabular}{|c|c|c|c|c|c|c|}
    \hline
        $\lambda$ & $L$ & $\rho$ & $\chi_1$ & $F$ & $\xi(L)$ & $\Upsilon$ \\     
        \hline
        0.00 & 320 & 0 & 693.32(77)  & 66.72(29) & 156.07(37)  & 0.6060(8)  \\
        0.01 & 320 & 0.0006(0) & 1003(1)     & 49.4(2)   & 223.69(62)  & 0.6367(4)  \\
        0.05 & 320 & 0.0049(0) & 1281.02(69) & 56.55(20) & 236.98(47)  & 0.6342(4)  \\
        0.10 & 320 & 0.0120(0) & 1406(1)     & 62.01(17) & 237.14(36)  & 0.6284(10) \\
        0.20 & 320 & 0.0293(0) & 1497(1)     & 67.2(2)   & 234.96(48)  & 0.6156(9)  \\
        0.30 & 320 & 0.0491(0) & 1492(1)     & 70.23(18) & 229.17(35)  & 0.5869(9)  \\
        0.40 & 320 & 0.0707(0) & 1403(1)     & 73.3(2)   & 216.97(33)  & 0.5349(7)  \\
        0.50 & 320 & 0.0940(0) & 1187(1)     & 81.10(22) & 188.08(32)  & 0.4158(6)  \\
        0.60 & 320 & 0.1190(0) & 697.9(7)    & 101.8(1)  & 123.23(13)  & 0.1633(6)  \\   
        0.65 & 320 & -         & 408.76(49)  & 106.03(14)& 86.06(9)    & -          \\
        0.66 & 320 & -         & 361.34(84)  & 104.91(25)& 79.63(16)   & -          \\
        0.70 & 320 & 0.1461(0) & 218.49(33)  & 93.91(7)  & 58.66(6)    & 0.0097(2)  \\
        0.72 & 320 & -         & 171.12(44)  & 85.57(17) & 50.92(9)    & -          \\
        0.75 & 320 & -         & 121.13(15)  & 72.89(7)  & 41.44(4)    & -          \\
        0.77 & 320 & -         & 97.81(19)   & 64.54(9)  & 36.57(10)   & -          \\
        0.80 & 320 & 0.1750(0) & 72.25(6)    & 53.09(4)  & 30.59(3)    & 0.0001(0)  \\
        0.84 & 320 & -         & 50.30(8)    & 40.76(6)  & 24.63(5)    & -          \\
        0.86 & 320 & -         & 42.33(8)    & 35.51(6)  & 22.32(4)    & -          \\
        0.89 & 320 & -         & 33.41(7)    & 29.19(5)  & 19.37(4)    & -          \\
        0.90 & 320 & -         & 31.05(7)    & 27.41(5)  & 18.55(5)    & -          \\
        1.00 & 320 & 0.2376(0) & 15.921(13)  & -         & -           & 0  \\
        \hline          
        0.00 & 640 & 0 & 1960(4) & 186.98(78) & 313.69(79)    & 0.6047(15)   \\
        0.01 & 640 & 0.0006(0) & 3345(6) & 151.9(8)   & 467(1)        & 0.6361(11)   \\
        0.05 & 640 & 0.0049(0) & 4319(3) & 186.63(68) & 479.29(97)    & 0.6328(9)    \\
        0.10 & 640 & 0.0120(0) & 4724(5) & 208.09(74) & 474.52(92)    & 0.6267(14)   \\
        0.20 & 640 & 0.0293(0) & 5002(4) & 224.3(9)   & 470(1)        & 0.6152(9)    \\
        0.30 & 640 & 0.0491(0) & 4919(5) & 234.20(80) & 455.58(91)    & 0.5818(12)   \\
        0.40 & 640 & 0.0707(0) & 4493(3) & 245.6(6)   & 423.63(64)    & 0.5098(8)    \\
        0.45 & 640 & -         & 4038(4) & 261.03(75) & 387.45(66)    & -             \\
        0.50 & 640 & 0.0940(0) & 3263(2) & 293.12(76) & 324.25(52)    & 0.3101(10)   \\
        0.55 & 640 & -         & 2087(3) & 340.57(53) & 230.68(26)    & -            \\
        0.60 & 640 & 0.1191(0) & 994(1)  & 332.0(4)   & 143.53(14)    & 0.0259(3)    \\   
        0.70 & 640 & 0.1461(0) & 221.05(31) & 165.49(17) & 59.02(6)   & 0   \\
        0.80 & 640 & 0.1750(0) & 72.36 (7)  & 66.37(6)   & 30.60(3)   & 0   \\
        0.84 & 640 & -         & 50.11(8)   & 47.33(7)   & 24.68(4)   & -           \\
        0.89 & 640 & -         & 33.44(6)   & 32.28(5)   & 19.38(4)   & -           \\
        0.90 & 640 & -         & 31.04(6)   & 30.04(5)   & 18.55(4)   & -           \\
        1.00 & 640 & 0.2377(0) & 15.905(12) &     -      &    -       & 0   \\
     \hline
    \end{tabular}
    \caption{Results for $\rho$, $\chi_1$, $F$, $\chi(L)$ and $\Upsilon$ for the fermionic $XY$ model, 
    $L=320$ and $L=640$.} 
    \label{Tab:X}
\end{table*}

\begin{table*}[!htbp]
\small
\renewcommand{\arraystretch}{1.2}
\setlength{\tabcolsep}{14pt}
    \centering
    \begin{tabular}{|c|c|c|c|c|c|c|}
    \hline
        $\lambda$ & $L$ & $\rho$ & $\chi_1$ & $F$ & $\xi(L)$ & $\Upsilon$ \\      
        \hline 
        0.00 & 1280 & 0 & 5531(13)   & 527(3) & 628(2)  & 0.6073(14)   \\
        0.01 & 1280 & 0.0006(0) & 11264(14)  & 486(3) & 959(4)  & 0.6363(16)   \\
        0.05 & 1280 & 0.0049(0) & 14466(15)  & 628(3) & 955(3)  & 0.6348(11)   \\
        0.10 & 1280 & 0.0120(0) & 15839(11)  & 696(3) & 949(2)  & 0.6281(12)   \\
        0.20 & 1280 & 0.0293(0) & 16702(15)  & 750(4) & 939(2)  & 0.6113(12)   \\
        0.25 & 1280 & -         & 16620(17)  & 773(2) & 922(2)  & -            \\
        0.30 & 1280 & 0.0491(0) & 16239(16)  & 782(3) & 905(1)  & 0.5772(8)    \\
        0.35 & 1280 & -         & 15466(12)  & 796(2) & 874(1)  & -            \\
        0.40 & 1280 & 0.0707(0) & 14072(14)  & 836(2) & 810(1)  & 0.4736(7)    \\
        0.45 & 1280 & -         & 11510(13)  & 934(2) & 686(1)  & -            \\
        0.50 & 1280 & 0.0940(0) & 7038(10)   & 1110(2)& 470(0)  & 0.1443(9)   \\
        0.60 & 1280 & 0.1191(0) & 1031(2)    & 683(1) & 145.51(24)   & 0.0004(1)  \\
        0.70 & 1280 & 0.1461(0) & 220.98(27) & 203.90(24) & 58.96(8) & 0 \\
        0.80 & 1280 & 0.1750(0) & 72.28(7)   & 70.68(7)   & 30.60(3) & 0 \\
        0.90 & 1280 & -         & 30.92(6)   & 30.67(6)   & 18.49(5) & -         \\
        1.00 & 1280 & 0.2377(0) & 15.924(11) & -          & -        & 0 \\ 
       \hline
        0.00 & 2560 & 0 & 15726(52)  & 1513(15)   & 1249(8)   & 0.6055(19)   \\
        0.01 & 2560 & 0.0006(0) & 37802(51)  & 1632(8)    & 1918(5)   & 0.6363(14)   \\
        0.05 & 2560 & 0.0049(0) & 48750(66)  & 2090(10)   & 1924(4)   & 0.6333(15)   \\
        0.10 & 2560 & 0.0120(0) & 53240(59)  & 2317(11)   & 1909(5)   & 0.6292(13)   \\
        0.20 & 2560 & 0.0293(0) & 55722(75)  & 2506(10)   & 1877(4)   & 0.6108(14)   \\
        0.25 & 2560 & -         & 55081(64)  & 2574(15)   & 1840(6)   & -            \\
        0.30 & 2560 & 0.0491(0) & 53348(50)  & 2616(8)    & 1793(3)   & 0.5685(9)    \\
        0.35 & 2560 & -         & 49684(68)  & 2693(11)   & 1702(4)   & -            \\
        0.40 & 2560 & 0.0707(0) & 42416(50)  & 2914(11)   & 1499(3)   & 0.4139(11)   \\
        0.45 & 2560 & -         & 27460(41)  & 3503(10)   & 1065(2)   & -            \\
        0.50 & 2560 & 0.0940(0) & 9492(16)   & 3496(3)    & 533(0)    & 0.0182(3)    \\
        0.60 & 2560 & 0.1191(0) & 1027(1)    & 912(1)     & 144.92(25) & 0  \\
        0.70 & 2560 & 0.1461(0) & 221.19(24) & 216.63(23) & 59.10(8)   & 0  \\
        0.80 & 2560 & 0.1750(0) & 72.26(8)   & 71.85(8)   & 30.53(3)   & 0  \\
        0.90 & 2560 & -         & 30.95(5)   & 30.88(5)   & 18.51(4)   & -           \\ 
        1.00 & 2560 & 0.2376(0) & 15.932(15) &  -         &  -         & 0  \\
        \hline          
        0.00 & 4000 & 0 & 30823(168)  & -  & -    & -      \\
        0.01 & 4000 & 0.0006(0) & 82478(194)  & -  & -    & -      \\
        0.05 & 4000 & 0.0049(0) & 106095(224) & -  & -    & -      \\
        0.1  & 4000 & 0.0120(0) & 116173(157) & -  & -    & -      \\
        0.2  & 4000 & 0.0293(0) & 121218(115) & -  & -    & -      \\
        0.3  & 4000 & 0.0491(0) & 114403(155) & -  & -    & -      \\
        0.4  & 4000 & 0.0707(0) & 82579(121)  & -  & -    & -      \\
        0.5  & 4000 & 0.0940(0) & 9743(19)    & -  & -    & -      \\
        0.6  & 4000 & 0.1191(0) & 1033(2)     & -  & -    & -      \\
        0.7  & 4000 & 0.1461(0) & 221.21(25)  & -  & -    & -      \\
        0.8  & 4000 & 0.1750(0) & 72.07(9)    & -  & -    & -      \\
        1.0  & 4000 & 0.2376(1) & 15.93(2)    & -  & -    & -      \\
     \hline
    \end{tabular}
    \caption{Results for $\rho$, $\chi_1$, $F$, $\chi(L)$ and $\Upsilon$ for the fermionic $XY$ model, 
    $L=1280$, $L=2560$ and $L=4000$.} 
    \label{Tab:XI}
\end{table*}

\begin{table*}[!htbp]
\small
\renewcommand{\arraystretch}{1.5}
\setlength{\tabcolsep}{4pt}
\begin{tabular}{c|ccc|ccc|ccc}
\TopRule
 & \multicolumn{3}{c|}{$L=40$} &
\multicolumn{3}{c|}{$L=80$} & \multicolumn{3}{c}{$L=160$} \\ 
\TopRule
$\beta$ & $\chi$ & $F$ & $\xi(L)$ & $\chi$ & $F$ & $\xi(L)$ & $\chi$ & $F$ & $\xi(L)$ \\
\MidRule
0.920 & 152.51(43) & 42.32(11) & 10.28(2) & 162.85(32) & 96.69(14)  & 10.54(2)  & 163.07(34)  & 139.21(29) & 10.54(3) \\
0.940 & 210.23(52) & 42.59(10) & 12.64(2) & 248.43(69) & 116.73(27) & 13.53(3)  & 250.93(76) & 195.02(47) & 13.64(4) \\
0.955 & 263.59(58) & 41.09(12) & 14.83(4) & 354(1)     & 130.41(32) & 16.68(4)  & 364(1)     & 252.31(63) & 16.94(5) \\
0.975 & 338.59(71) & 38.39(14) & 17.82(5) & 581(1)     & 139.79(44) & 22.62(6)  & 647(2)     & 347(1)     & 23.67(8)\\
0.980 & 359.02(60) & 37.72(12) & 18.60(3) & 658(2)     & 140.37(34) & 24.46(6)  & 759(2)     & 371.54(87) & 26.01(6) \\
0.985 & 378.60(77) & 36.66(17) & 19.46(5) & 735(2)     & 138.42(44) & 26.45(7)  & 895(3)     & 397.05(99) & 28.51(9) \\
0.990 & 395.35(67) & 36.09(15) & 20.11(6) & 820(1)     & 137.25(35) & 28.41(6)  & 1069(4)    & 418(1)     & 31.79(10) \\
0.995 & 411.92(88) & 35.16(15) & 20.86(6) & 911(2)     & 134.31(41) & 30.62(7)  & 1276(4)    & 437(1)     & 35.32(10) \\
1.000 & 428.89(71) & 34.46(16) & 21.56(6) & 999(2)     & 130.56(51) & 32.85(8)  & 1532(4)    & 454(2)     & 39.25(14) \\
1.010 & 459.68(93) & 33.00(14) & 22.91(7) & 1181(3)    & 123.96(46) & 37.18(9)  & 2158(7)    & 461(1)     & 48.89(13) \\
1.015 & 474.58(62) & 32.37(21) & 23.55(8) & 1270(3)    & 121.18(67) & 39.21(14) & 2539(9)    & 460(1)     & 54.16(14) \\
1.024 & 500.46(56) & 31.71(12) & 24.50(5) & 1416(2)    & 114.09(60) & 43.02(12) & 3280(12)   & 437(1)     & 64.93(23) \\
1.026 & 505.80(56) & 31.22(11) & 24.85(5) & 1443(4)    & 112.83(61) & 43.73(16) & 3454(8)    & 433(2)     & 67.23(16) \\
1.030 & 515.03(68) & 30.74(13) & 25.29(7) & 1499(2)    & 110.99(46) & 45.03(10) & 3771(9)    & 420(2)     & 71.91(20) \\
1.035 & 526(1)     & 30.14(15) & 25.86(9) & 1569(3)    & 108.42(47) & 46.74(11) & 4162(8)    & 402(2)     & 77.85(24) \\
1.064 & 586.41(74) & 28.22(14) & 28.34(9) & 1891(3)    & 97.25(74) & 54.70(24) & 5935(10)    & 342(3)     & 103.02(55) \\ 
1.066 & 591.16(64) & 27.83(21) & 28.67(12)& 1907(3)    & 97.54(53) & 54.86(17) & 6034(10)    & 332(2)     & 105.48(45) \\
1.070 & 597.61(69) & 27.64(18) & 28.94(10)& 1944(2)    & 97.22(56) & 55.50(17) & 6221(13)    & 334(2)     & 106.96(36) \\
1.078 & 609.68(72) & 27.84(11) & 29.13(7)  & 2007(2)   & 94.37(52)  & 57.33(17) & 6562(10)   & 323(2)     & 111.99(44) \\
1.084 & 620.71(96) & 27.50(10) & 29.60(6)  & 2050(4)   & 93.65(67)  & 58.22(23) & 6731(12)   & 322(2)     & 113.60(45) \\
1.104 & 648.64(76) & 26.54(16) & 30.86(10) & 2184(3)   & 91.11(64)  & 61.04(25) & 7343(11)   & 313(2)     & 120.78(42) \\
1.111 & 659(1)     & 26.40(22) & 31.21(15) & 2233(4)   & 90.99(89)  & 61.79(33) & 7531(15)   & 310(3)     & 122.85(76) \\
\BotRule
\end{tabular}
\caption{Results for $\chi$, $F$ and $\chi(L)$ for the bosonic $XY$ model, $L=40$, $L=80$, and $L=160$.}
\label{Tab:XII}
\end{table*}

\begin{table*}[!htbp]
\small
\renewcommand{\arraystretch}{1.5}
\setlength{\tabcolsep}{4pt}
\begin{tabular}{c|ccc|ccc|ccc}
\TopRule
 & \multicolumn{3}{c|}{$L=320$} &
\multicolumn{3}{c|}{$L=640$} & \multicolumn{3}{c}{$L=1280$} \\ 
\TopRule
$\beta$ & $\chi$ & $F$ & $\xi(L)$ & $\chi$ & $F$ & $\xi(L)$ & $\chi$ & $F$ & $\xi(L)$ \\
\MidRule
0.920 & 162.08(49) & 155.43(45) & 10.53(4) & 162.71(42) & 160.99(41) & 10.54(3)   & 161(2)   & 160(2)   & 10.53(6) \\
0.940 & 251.67(39) & 234.84(36) & 13.63(4) & 251.09(66) & 246.70(65) & 13.59(3)   & 249(2)   & 248(2)   & 13.59(6) \\
0.955 & 364(1)     & 328(1)   & 16.88(5)   & 363(1)     & 353(1)     & 16.81(7)   & 360(3)   & 358(2)   & 16.77(5) \\
0.975 & 645(2)     & 531(2)   & 23.59(8)   & 652(2)     & 618(1)     & 23.66(8)   & 647(2)   & 639(2)   & 23.44(8)\\
0.980 & 767(2)     & 609(1)   & 26.03(7)   & 765(2)     & 718(2)     & 26.16(9)   & 766(3)   & 753(3)   & 26.13(10) \\
0.985 & 909(3)     & 691(1)   & 28.61(12)  & 906(2)     & 840(2)     & 28.63(9)   & 909(3)   & 891(3)   & 28.72(10) \\
0.990 & 1086(3)    & 780(2)   & 31.91(8)   & 1083(3)    & 986(2)     & 31.85(14)  & 1091(4)  & 1065(4)  & 31.76(14) \\
0.995 & 1308(5)    & 876(3)   & 35.74(11)  & 1321(6)    & 1176(6)    & 35.78(14)  & 1313(5)  & 1274(5)  & 35.52(14) \\
1.000 & 1609(5)    & 994(2)   & 40.07(10)  & 1612(6)    & 1396(4)    & 40.12(17)  & 1605(5)  & 1546(4)  & 39.91(14) \\
1.010 & 2479(10)   & 1227(3)  & 51.44(17)  & 2510(8)    & 1993(5)    & 51.90(19)  & 2498(8)  & 2346(7)  & 51.85(28) \\
1.015 & 3182(13)   & 1354(4)  & 59.16(16)  & 3220(14)   & 2392(8)    & 59.93(27)  & 3200(13) & 2943(11) & 60.15(24) \\
1.024 & 493116)    & 1494(5)  & 77.24(19)  & 5204(14)   & 3241(10)   & 79.26(32)  & 5257(30)    & 4564(22) & 79.41(36) \\   
1.026 & 5465(18)   & 1513(5)  & 82.30(29)  & 5834(28)   & 3455(11)   & 84.52(36)  & 5878(19)    & 5013(17) & 84.64(24) \\
1.030 & 6660(22)   & 1549(6)  & 92.52(27)  & 7439(36)   & 3893(11)   & 97.23(43)  & 7499(34)    & 6092(23) & 97.89(36) \\
1.035 & 8342(34)   & 1520(8)  & 107.90(34) & 10258(48)  & 4422(17)   & 117.01(49) & 10447(36)   & 7807(23) & 118.46(38) \\
1.064 & 18032(36)  & 1187(8)  & 191.87(73) & 49738(162) & 4371(35)   & 328(1)     & 108892(596) & 16547(126) & 481(2) \\
1.066 & 18469(44)  & 1185(8)  & 194.64(80) & 52457(122) & 4299(30)   & 341(1)     & 124091(744) & 16207(149) & 526(4) \\
1.070 & 19466(43)  & 1176(7)  & 200.89(66) & 57232(180) & 4110(25)   & 366(2)     & 151071(649) & 15152(146) & 612(4) \\
1.078 & 20924(52)  & 1112(9)  & 214.95(95) & 65456(205) & 3877(31)   & 406(2)     & 197058(841) & 13360(128) & 755(4) \\
1.084 & 21967(47)  & 1103(9)  & 221(1)     & 70376(200) & 3802(23)   & 426(2)     & 221594(633) & 12822(111) & 822(4) \\
1.104 & 24538(58)  & 1048(7)  & 241(1)     & 82560(241) & 3582(29)   & 478(2)     & 274700(716) & 12070(122) & 950(6) \\
1.111 & 25340(56)  & 1062(10) & 243(1)     & 85517(263) & 3617(41)   & 485(3)     & 288454(1069) & 12234(164)& 968(7) \\
\BotRule
\end{tabular}
\caption{Results for $\chi$, $F$ and $\chi(L)$ for the bosonic $XY$ model, $L=320$, $L=640$ and $L=1280$.}
\label{Tab:XIII}
\end{table*}


\end{document}